\documentclass[prd,12pt]{article}
\pdfoutput=1 

\usepackage{tikz}
\usepackage{amsmath,amssymb,graphicx,multirow,xspace,slashed,array,booktabs}
\usepackage[colorlinks=true,urlcolor=blue,anchorcolor=blue,citecolor=blue,filecolor=blue,linkcolor=blue,menucolor=blue,pagecolor=blue]{hyperref}
\usepackage[compress,numbers]{natbib}
\usepackage{subfigure, placeins}
\usepackage{booktabs}
\usepackage{blindtext, rotating}
\usepackage{afterpage}
\usepackage{enumitem}
\usepackage{ marvosym }
\usepackage{authblk} 
\usepackage{verbatim}
\usepackage{soul} 
\usepackage[normalem]{ulem}
\usepackage{pifont}
\usepackage{lscape}
\usepackage{threeparttable}
\usepackage{threeparttablex}
\usepackage[export]{adjustbox}

\usepackage[most]{tcolorbox}

\usepackage{cellspace}
\usepackage{floatrow}
\newfloatcommand{capbtabbox}{table}[][\FBwidth]

\usepackage[font=footnotesize,labelfont=bf]{caption}

\usepackage{lineno}

\allowdisplaybreaks

\long\def\symbolfootnote[#1]#2{\begingroup%
\def\thefootnote{\fnsymbol{footnote}}\footnote[#1]{#2}\endgroup}

\allowdisplaybreaks

\newcommand{\newc}{\newcommand}
\newc{\gsim}{\lower.7ex\hbox{$\;\stackrel{\textstyle>}{\sim}\;$}}
\newc{\lsim}{\lower.7ex\hbox{$\;\stackrel{\textstyle<}{\sim}\;$}}
\newc{\gev}{\,{\rm GeV}}
\newc{\mev}{\,{\rm MeV}}
\newc{\ev}{\,{\rm eV}}
\newc{\kev}{\,{\rm keV}}
\newc{\tev}{\,{\rm TeV}}
\newc{\MHT}{$H_T^{\text{miss}}$}
\newc{\MET}{$\slashed{E}_T$}
\newc{\MTT}{$M_{T2}$}

\newc{\mz}{M_Z}
\newc{\mpl}{M_*}
\newc{\mw}{m_{\rm weak}}
\newc{\nr}[1]{N^c_R{}_{#1}}

\def\beq{\begin{equation}}
\def\eeq{\end{equation}}
\newcommand{\bea}{\begin{eqnarray}\begin{aligned}}
\newcommand{\eea}{\end{aligned}\end{eqnarray}}
\def\bitem{\begin{itemize}}
\def\eitem{\end{itemize}}

\usepackage{xcolor}
\definecolor{cambridgeblue}{rgb}{0.64, 0.76, 0.68}
\definecolor{darkraspberry}{rgb}{0.53, 0.15, 0.34}

\begin{document}

\baselineskip 0.6cm

\begin{titlepage}

\thispagestyle{empty}

\begin{center}

\vskip 0.5cm


{\Huge \bf Boosted $W/Z$ Tagging with }\vskip0.3cm
{\Huge \bf Jet Charge and Deep Learning }

\vskip 0.1cm

\vskip 1.0cm
{\large Yu-Chen Janice Chen$^1$, Cheng-Wei Chiang$^{1,2,3}$,}
\vskip0.2cm
{\large  Giovanna Cottin$^{1,4,5}$ and David Shih$^6$ }
\vskip 1.0cm

{\it $^1$ Department of Physics, National Taiwan University, Taipei 10617, Taiwan}\\
{\it $^2$ Institute of Physics, Academia Sinica, Taipei 11529, Taiwan}\\
{\it $^3$ Department of Physics and Center of High Energy and High Field Physics, National Central University, Chungli 32001, Taiwan}\\
{\it $^4$ Instituto de F\'isica, Pontificia Universidad Cat\'olica de Chile, Avenida Vicu\~{n}a Mackenna 4860, Santiago, Chile}\\
{\it $^5$ Departamento de Ciencias, Facultad de Artes Liberales, Universidad Adolfo Ib\'a\~{n}ez, Diagonal Las Torres 2640, Santiago, Chile}\\
{\it $^6$ NHETC, Dept. of Physics and Astronomy, Rutgers, The State University of NJ Piscataway, NJ 08854, USA}

\vskip 0.8cm

\end{center}

\vskip 0.8cm

\begin{abstract}
We demonstrate that the classification of boosted, hadronically-decaying weak gauge bosons can be significantly improved over traditional cut-based and BDT-based methods using deep learning and the jet charge variable. We construct binary taggers for $W^+$ vs.\ $W^-$ and $Z$ vs.\ $W$ discrimination, as well as an overall ternary classifier for $W^+$/$W^-$/$Z$ discrimination. Besides a simple convolutional neural network (CNN), we also explore a composite of two CNNs, with different numbers of layers in the jet $p_{T}$ and jet charge channels. We find that this novel structure boosts the performance particularly when considering the $Z$ boson as signal. The methods presented here can enhance the physics potential in SM measurements and searches for new physics that are sensitive to the electric charge of weak gauge bosons. 
\end{abstract}

\flushbottom

\end{titlepage}

\section{Introduction} 

Boosted heavy resonances play a central role in the study of physics at the Large Hadron Collider (LHC). These include both Standard Model (SM) particles such as $W$'s, $Z$'s, tops and Higgses, as well as hypothetical new physics (NP) particles such as $Z'$'s. The decay products of the boosted heavy resonance are typically collimated into a single ``fat jet" with nontrivial internal substructure. A vast amount of effort has been devoted to the important problem of ``tagging" ({\it i.e.}, identifying and classifying) boosted resonances through the understanding of jet substructure. (For recent reviews and original references, see e.g.~\cite{Larkoski:2017jix,Asquith:2018igt,Marzani:2019hun}.) 

Recently, there has been enormous interest in the application of modern deep learning techniques to
boosted resonance tagging~\cite{Almeida:2015jua, deOliveira:2015xxd, Baldi:2016fql, Guest:2016iqz, Komiske:2016rsd,2017JHEP...05..006K, Louppe:2017ipp, Pearkes:2017hku, 2018ScPP....5...28B,  Cheng:2017rdo, Egan:2017ojy, ATL-PHYS-PUB-2017-004, Macaluso:2018tck, Datta:2019ndh, Qu:2019gqs, Kasieczka:2019dbj,  Dillon:2019cqt, Bhattacherjee:2019fpt,Diefenbacher:2019ezd, Moreno:2019bmu,CMS-PAS-JME-18-002}. By enabling the use of high-dimensional, low-level inputs (such as jet constituents), deep learning automates the process of feature engineering. Many works have demonstrated the enormous potential of deep learning to construct extremely powerful taggers, vastly improving on previous methods.   

So far, most of the attention has focused on distinguishing various boosted resonances from QCD background in a binary classification task. Less attention has been paid to multi-class classification, {\it i.e.}, a tagger that would categorize jets in a multitude of possibilities. (Notable exceptions include refs.~\cite{Moreno:2019bmu,CMS-PAS-JME-18-002}.) In this work, we will examine an important multi-class classification task: distinguishing $W^+$, $W^-$ and $Z$ bosons.
Having a $W/Z$ classifier that can also recognize charge could have many interesting applications. For instance, one could use such a classifier to measure charge asymmetries and same-sign diboson production at the LHC. Or there are many potential applications to NP scenarios, such as the reconstruction of doubly-charged Higgs bosons from its like-sign diboson decay in models with an extended scalar sector.

Since we are interested in distinguishing $W^+$ and $W^-$ bosons from each other, a key element in our work will be the jet charge observable $\mathcal{Q}_{\kappa}$.  It was first introduced in ref.~\cite{Field:1977fa} and its theoretical potential was discussed further in ref.~\cite{Krohn:2012fg}. 
Such an observable has also been measured at the LHC~\cite{Aad:2015cua,Sirunyan:2017tyr}. When used in conjunction with other quantities, such as the invariant mass ${\mathcal{M}}$, it can help distinguish between hadronically decaying $W$ from $Z$ bosons~\cite{Aad:2015eax}. Moreover, its performance as a charge-tagger was assessed in ref.~\cite{ATLAS-CONF-2013-086} for jets produced in semileptonic $t\bar{t}$, $W+$jets and dijet processes. Most recently the authors of ref.~\cite{Fraser:2018ieu} incorporated jet charge into various machine learning quark/gluon taggers, including BDTs, convolutional neural networks (CNNs), and recurrent neural networks (RNNs). They showed that including jet charge in the input channels improved quark/gluon discrimination and up vs.\ down quark discrimination. 

In our study of $W^+/W^-/Z$ tagging, we will compare a number of techniques, from simple cut-based methods, to BDTs, to deep learning methods based on CNNs and jet images. As in ref.~\cite{Fraser:2018ieu}, we will include jet charge as one of the input channels and examine the gain in performance from including this additional input. We will go beyond refs.~\cite{Aad:2015eax,ATL-PHYS-PUB-2017-004,Fraser:2018ieu} and construct a ternary classifier that can discriminate among $W^+$, $W^{-}$, and $Z$, depending on the physics process of interest. We will study the overall performance of our ternary tagger as well as its specialization to binary classification. For the latter we will compare its performance to specifically trained binary classifiers and show that the ternary classifier reproduces their performance, and in this sense it is optimal. Overall, we will demonstrate that deep learning with jet charge offers a significant boost in performance, around $\sim$30-40\% improvement in background rejection rate at fixed signal efficiency.

In addition to a simple CNN, we will also develop a novel composite algorithm consisting of two CNNs, one for each of the $\mathcal{Q}_{\kappa}$ and $p_{T}$ channels, combined in a merge layer, which we refer to as CNN$^{2}$. This allows us to separately optimize the hyperparameters of the CNNs for the two input channels. We show that this new CNN$^{2}$ architecture further boosts the performance for most combinations.

The rest of the paper is organized as follows. In Sec.~\ref{sec:observables} we describe the jet samples and jet images used in this study, and review the definition of the jet charge variable.
 In Sec.~\ref{sec:jetChargeTagger}, we describe the different taggers studied in this paper. These include cut-based and BDT taggers used as baselines for comparison, as well as two different taggers based on convolutional neural networks.
We show results for a binary $W^{-}/W^{+}$ classification problem in Sec.~\ref{sec:binary_wpwn} and compare our performance with the recent work in ref.~\cite{Fraser:2018ieu}. In Sec.~\ref{sec:binary_wz}, we discuss the $Z/W^{+}$ discrimination problem, focusing on the benefit from including jet charge, and compare our performance with the ATLAS boson tagger in ref.~\cite{Aad:2015eax}. Finally in Sec.~\ref{sec:3to2}, we extend our results to the full ternary $Z/W^{+}/W^{\text{-}}$ classification problem, and comment on the reduction from a three-class tagger to a two-class one. In Sec.~\ref{sec:visualization}, we attempt to shed some light on what the deep neural networks learned
We summarize our findings and conclude in Sec.~\ref{sec:summary}. 

\section{Jet samples and inputs}
\label{sec:observables}

For this study, we use \textsc{MadGraph5v2.6.1}~\cite{Alwall:2014hca} at leading order to simulate events at the 13~TeV LHC for VBF production of doubly charged Higgses $H^{\pm\pm}_{5}$ and heavy neutral Higgses $H_5$, with decays $H^{\pm\pm}_{5}\to W^{\pm}W^{\pm}\to jjjj$ and $H_{5} \to ZZ\to jjjj$ respectively. We take the spectrum of the exotic Higgs bosons in the Georgi-Machacek model generated by~\textsc{GMCalc}~\cite{Hartling:2014xma} as the input. For simplicity, we fix $m_{{H}_{5}} = 800$~GeV, so that the $p_T$ of each vector boson is typically  $\sim 400$~GeV.\footnote{We have also studied the scenario where $m_{H_5} = 2$~TeV that leads to $p_T$ of each boson around 1~TeV, and found all results qualitatively the same.}
All events are further processed in~\textsc{PYTHIA 8.2.19}~\cite{Sjostrand:2014zea} for showering and hadronization and passed onto \textsc{DELPHES}~3.4.1~\cite{deFavereau:2013fsa} with the default CMS card for detector simulation.\footnote{A dataset based on \textsc{HERWIG} showering and hadronization is also generated for the purpose of checking the reliability of the deep learning jet-tagging results. Details are given in appendix~\ref{sec:herwig}. } Jets are reconstructed with \textsc{FastJet}~3.1.3~\cite{Cacciari:2011ma} using the anti-$k_{T}$ clustering algorithm \cite{Cacciari:2008gp} with the jet radius parameter $R =0.7$. This jet radius is appropriate to the $p_T$ range indicated above, as it will capture most of the hadronic $W$ and $Z$ decay products.

The jets selected in this work are required to satisfy $|\eta|\leq 1$. 
The jets must also be truth-matched to a $W/Z$ boson by requiring their distance between the jet and the vector boson, known from truth Monte Carlo, in the $(\eta,\phi)$ plane be less than $0.1$. A summary detailing the full set of selections for our jet sample is given in table~\ref{tab:selections}.  The jet sample sizes of the training and testing sets for our taggers are summarized in table~\ref{tab:JetSamples}.

In what follows, we consider a number of inputs in the construction of our BDT and CNN taggers. 

\begin{table}[t]
\begin{center}
\begin{tabular*}{0.6\textwidth}{c|c}
\bottomrule 
           & $p_{T} \in (350,450)$ GeV, $|\eta|\leq 1$  \\
Jet sample &jets with anti-$k_{T}$ and $R=0.7$ \\ 
           &   $V$-$V$ merging : $\Delta R(V_{1},V_{2}) < 0.6$ \\ 

  &   $V$-jet matching : $\Delta R(V,j) < 0.1$        \\
\toprule 
\end{tabular*} 
\caption{Selections imposed on the jet sample used in our analyses.}
\label{tab:selections}
\end{center}
\end{table}

\begin{table}[h]
\begin{center}
\begin{tabular*}{0.75\textwidth}{c|c|c}
\bottomrule
						 & \hspace{0.2cm} Training set\hspace{0.2cm}  & \hspace{0.2cm} Testing set \hspace{0.2cm} \\
\hline 
\hspace{0.2cm} Jet sample size    & \hspace{0.2cm} $188k+198k+175k$ \hspace{0.2cm} & $38k+40k+35k$ \hspace{0.2cm}\\
\toprule
\end{tabular*} 
\caption{Summary of the jet sample sizes used for training and testing, after the selections in table~\ref{tab:selections}. The entries in the sum correspond to the ($W^+, W^-, Z$) samples, respectively. In the training stage, the training set is further divided into two subsets: $9/10$ is the actual training set, and $1/10$ serves as the validation set. All the ROC and SIC curves shown in this paper are evaluated using the testing sets described here.}
\label{tab:JetSamples}
\end{center}
\end{table}

\subsection{Jet mass and jet charge ($\mathcal{M}$, $\mathcal{Q}_{\kappa}$) observables}

For the problem at hand, there are two obvious high-level observables: jet invariant mass and jet charge.  The jet invariant mass $\cal M$ is defined as
\begin{equation}
{\cal M}^2
=
\left( \sum_{i \in J} E_i \right)^2 - \left( \sum_{i \in J} {\bf p}_i \right)^2
~.
\end{equation}
The sum runs over all constituents $i$ inside the jet $J$ ({\it i.e.}, all tracks and calorimeter towers in the jet) with 4-momentum $(E_i,{\bf p}_i)$ and $p_{T}>500$~MeV.

\begin{figure}[ht]
\centering
\includegraphics[width=0.45\textwidth,angle=0]{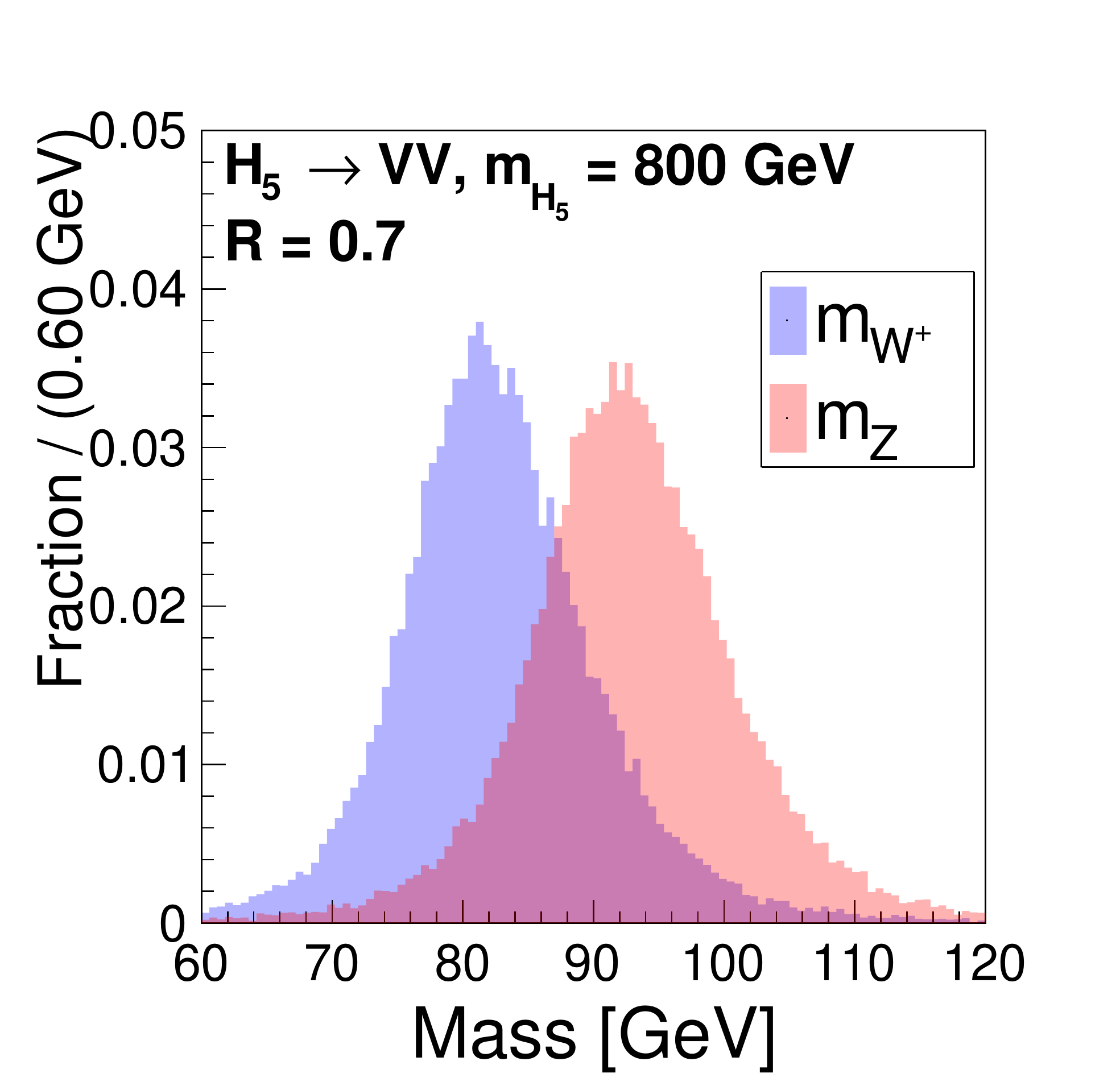}
\caption{Reconstructed jet mass of $W$ and $Z$ samples.}
\label{fig:jetMass}
\end{figure}

Figure~\ref{fig:jetMass} shows the reconstructed jet mass after all selections in table~\ref{tab:selections}.  The broader widths in the mass distribution originate from a combination of showering, hadronization, jet clustering and detector effects.

It has been known for some time~\cite{Field:1977fa} that a useful observable for distinguishing jets initiated by particles of different charges is the jet charge:
\begin{equation}
\mathcal{Q}_{\kappa} = \frac{1}{(p_{T, J})^{\kappa}} \sum_{i \in J} q_{i}\times (p^{i}_{T})^{\kappa}
~,
\label{eq:jetcharge}
\end{equation}
where $q_{i}$ corresponds to the integer charge of the jet constituent in units of the proton charge, and $\kappa$ is a free parameter. The $\mathcal{Q}_{\kappa}$ observable is computed in this $p_{T}$-weighted scheme to minimize mis-measurements from low-$p_{T}$ particles.

\begin{figure}[ht]
\centering
\includegraphics[width=\textwidth,angle=0]{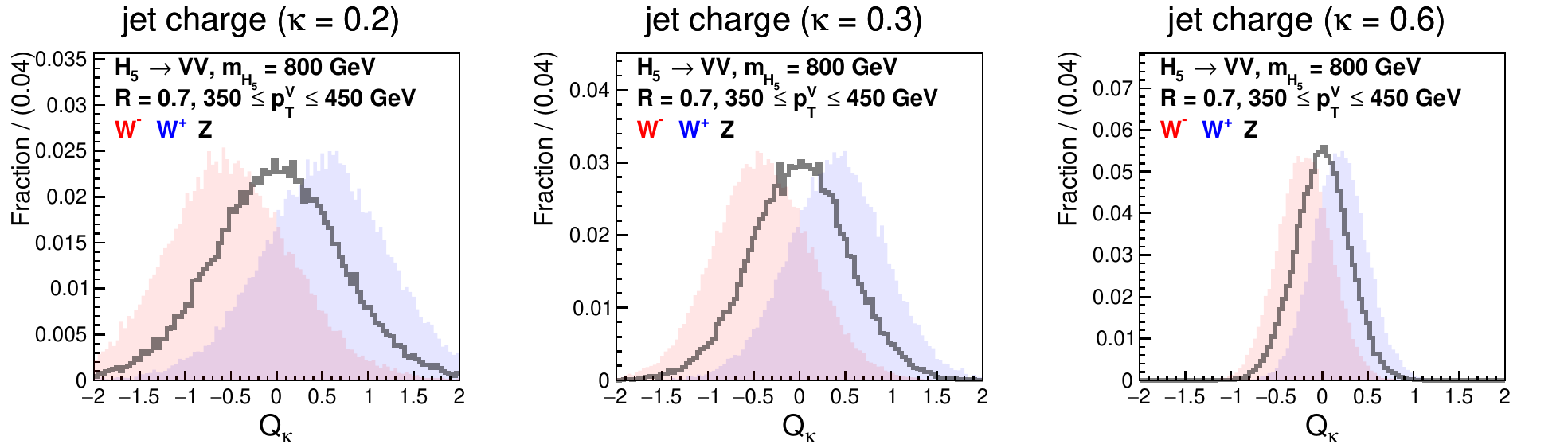}
\caption{$\mathcal{Q}_{\kappa}$ distributions for the three samples under study.  Representative $\kappa$ values are shown.}
\label{fig:jetCharge}
\end{figure}

Figure~\ref{fig:jetCharge} shows the $\mathcal{Q}_{\kappa}$ distributions for jets coming from the $W^{+}$, $W^{-}$ and $Z$ samples. Distributions are shown for different $\kappa$ values. The choice of $\kappa$  together with the $p_{T}$ range of the vector bosons will affect the tagging performance. 

\subsection{Jet images}
\label{sec:JetImages}

As described in the introduction, the deep learning based taggers studied in this work are based on jet images and convolutional neural networks. In this work, our jet images are made from jets reconstructed in a $\Delta\eta=\Delta\phi=1.6$ box, with $75\times 75$ pixels. 
 This choice yields a resolution consistent with that of the CMS ECal (see table~\ref{tab:Networks}). The input variables or channels are now $\mathcal{Q}_{\kappa}$ and $p_{T}$ per pixel.  Therefore, the sum in the $\mathcal{Q}_{\kappa}$ definition in Eq.~\eqref{eq:jetcharge} in this case goes over all jet constituents in each pixel. 

\begin{figure}[h]
\centering
\includegraphics[width=0.48\textwidth,angle=0]{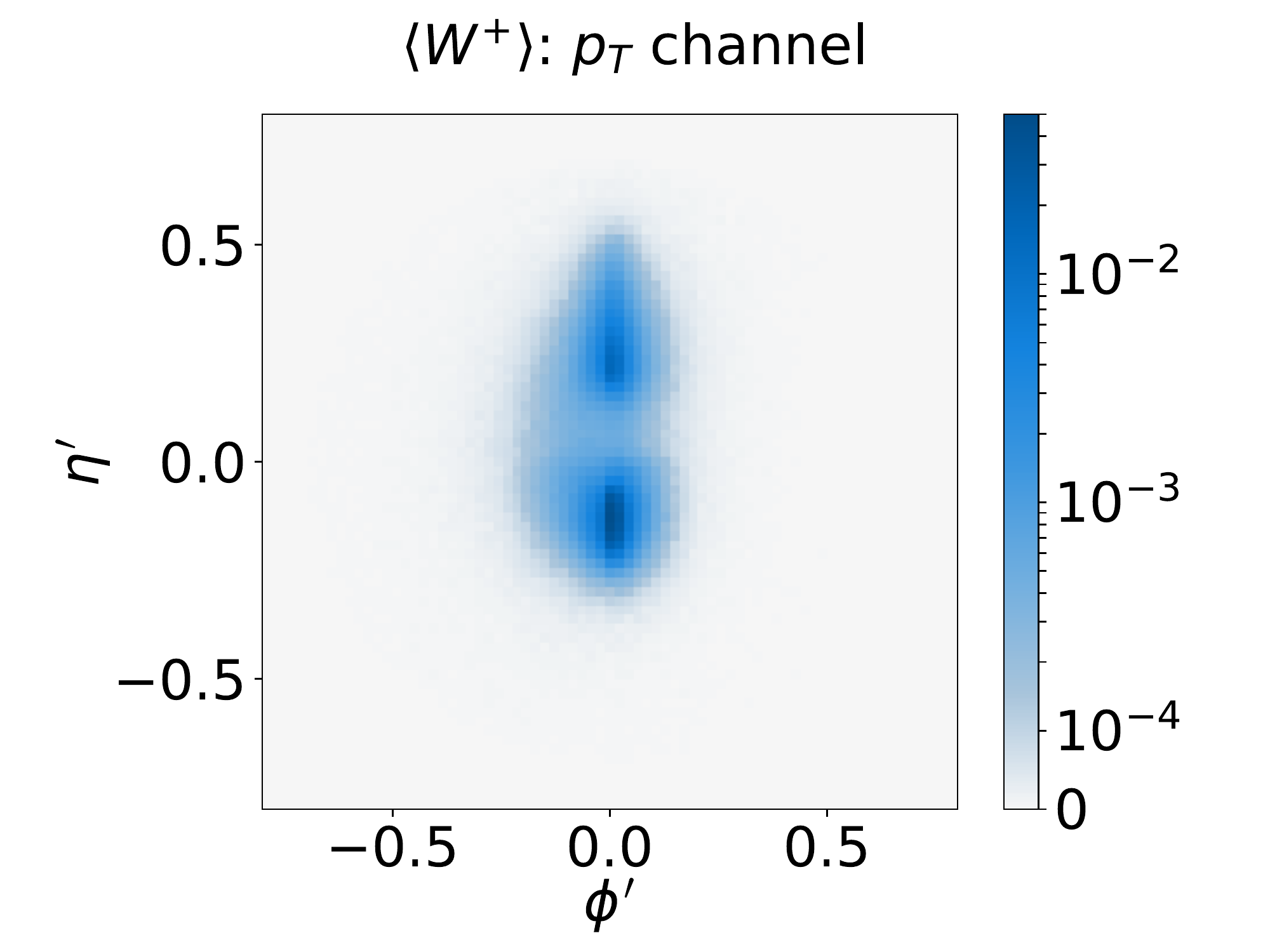}
\includegraphics[width=0.48\textwidth,angle=0]{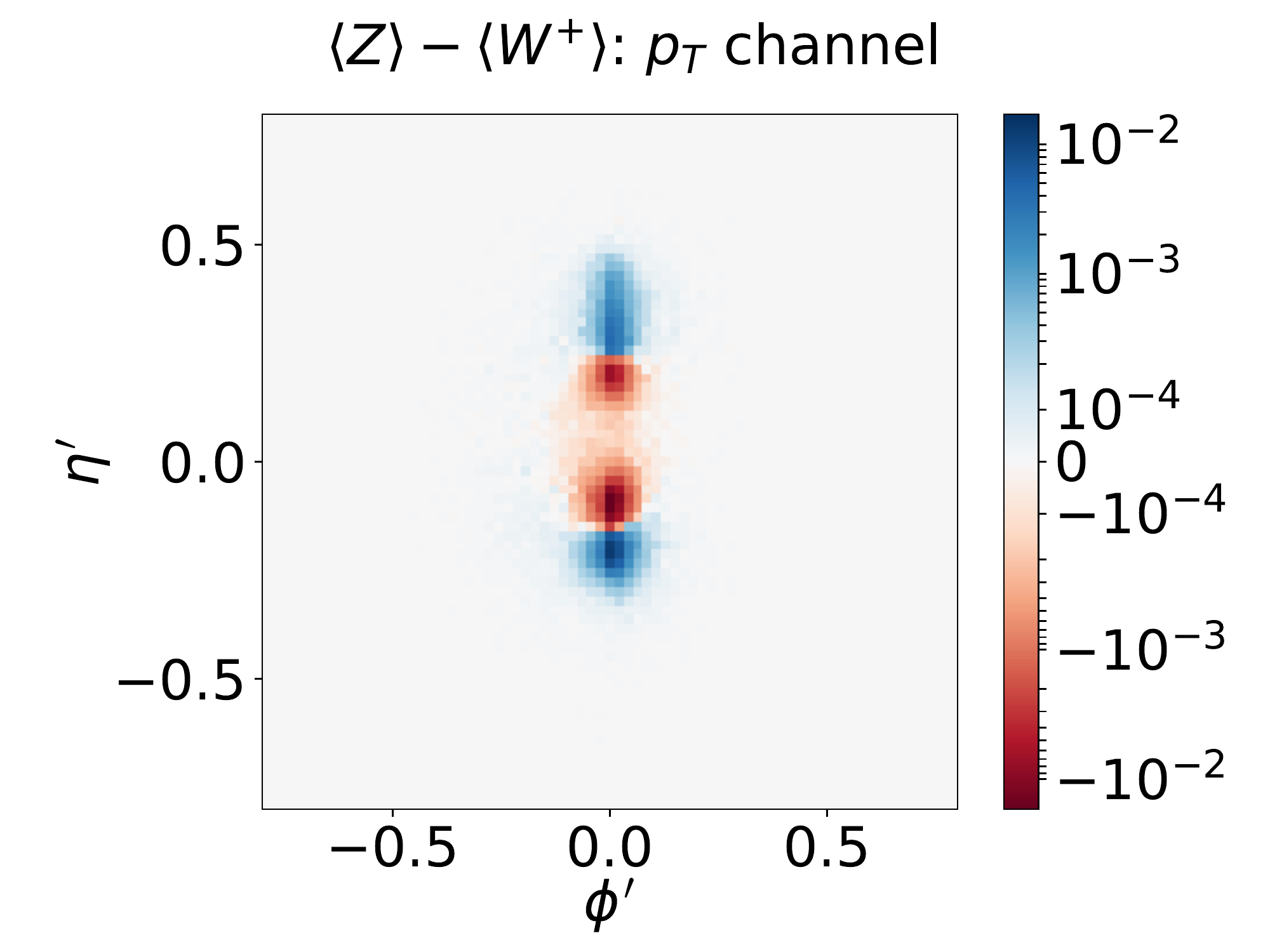}
\caption{The left plot shows the average of $W^{+}$ jet images in the $p_{T}$ channel using the preprocessed testing set sample. The right plot shows the difference between $Z$ and $W^{+}$ average jet images in the $p_{T}$ channel.
}
\label{fig:jetImages}
\end{figure}

\begin{figure}[h]
\centering
\includegraphics[width=0.48\textwidth,angle=0]{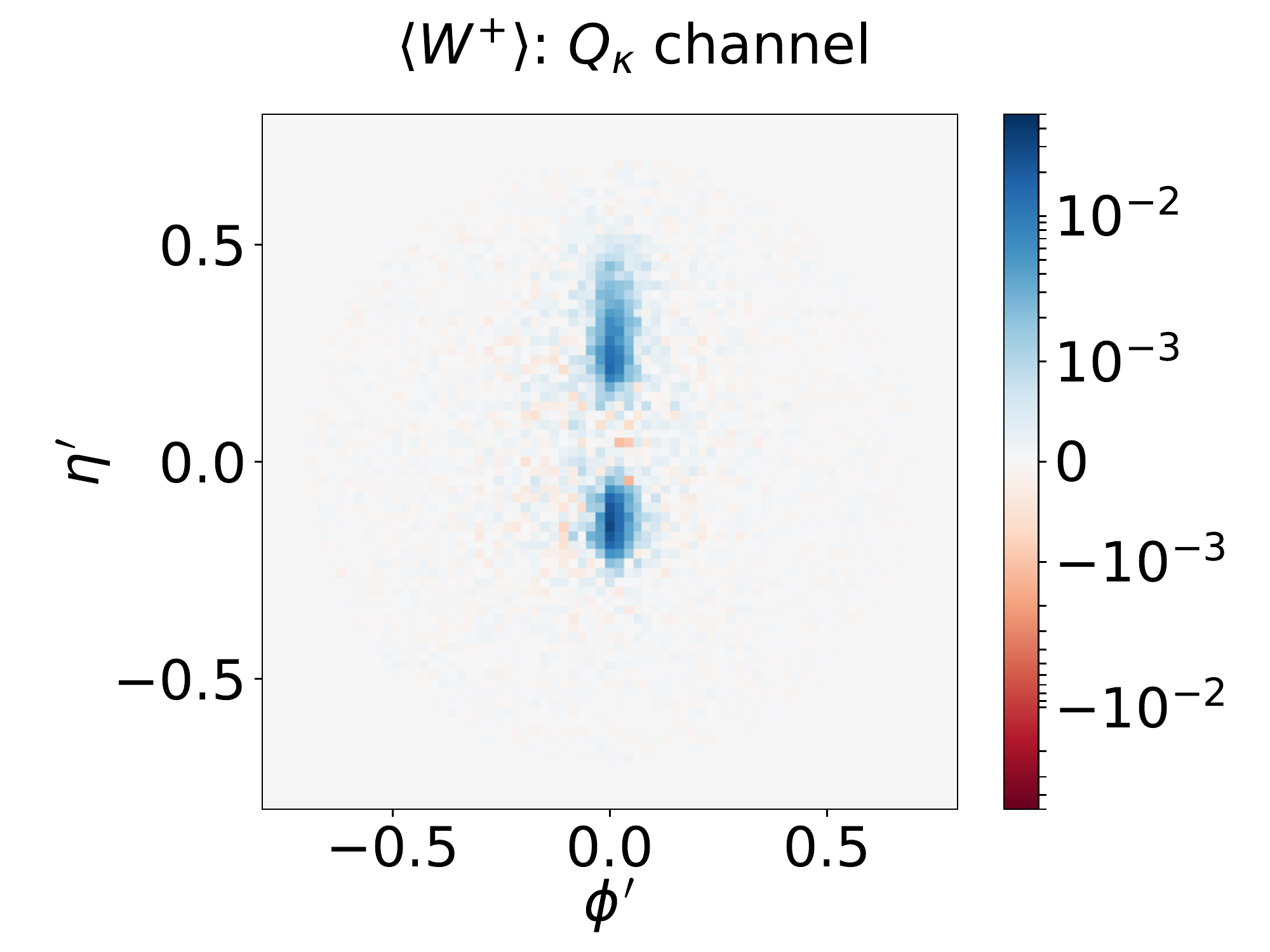}
\includegraphics[width=0.48\textwidth,angle=0]{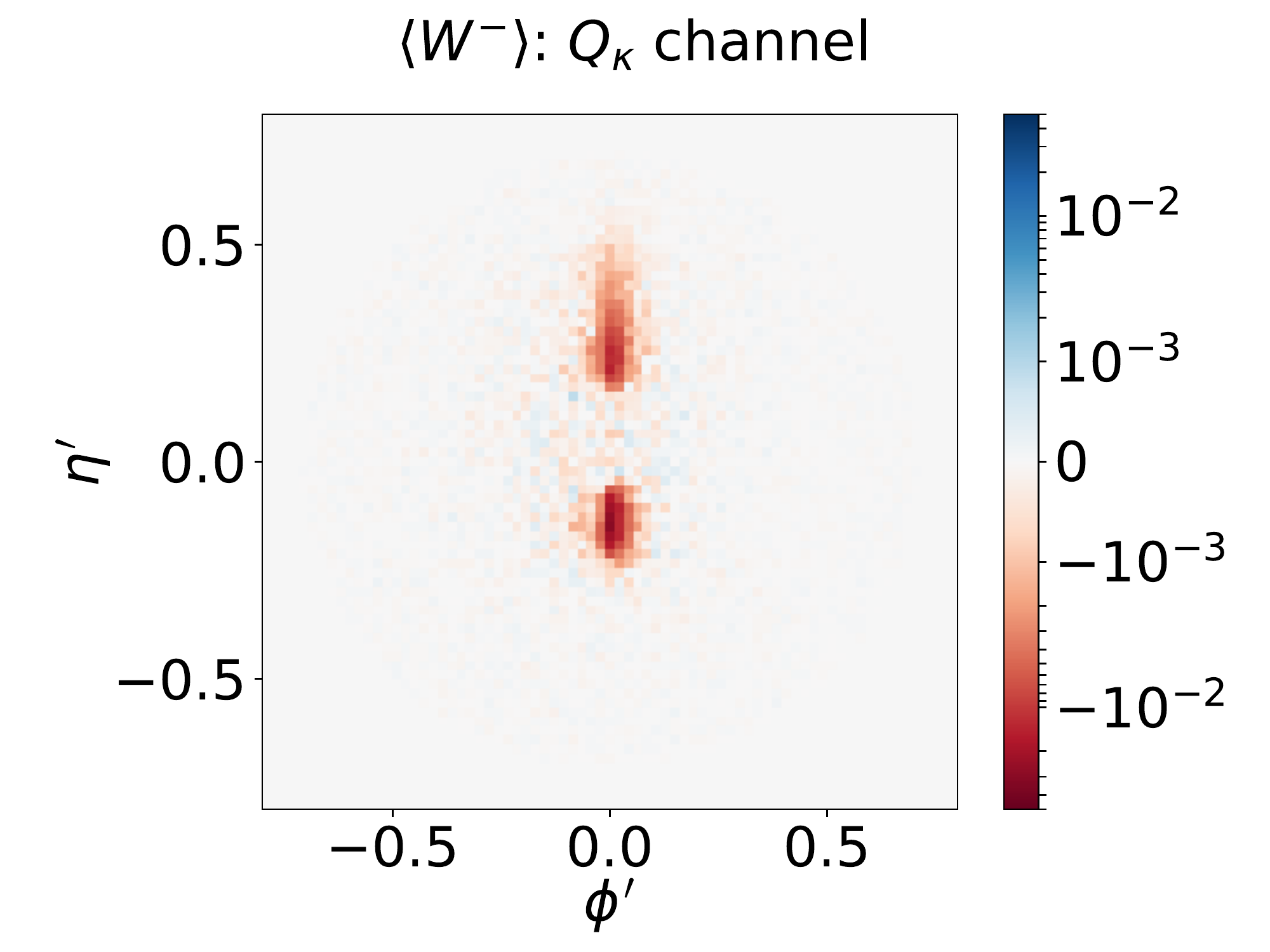}
\includegraphics[width=0.48\textwidth,angle=0]{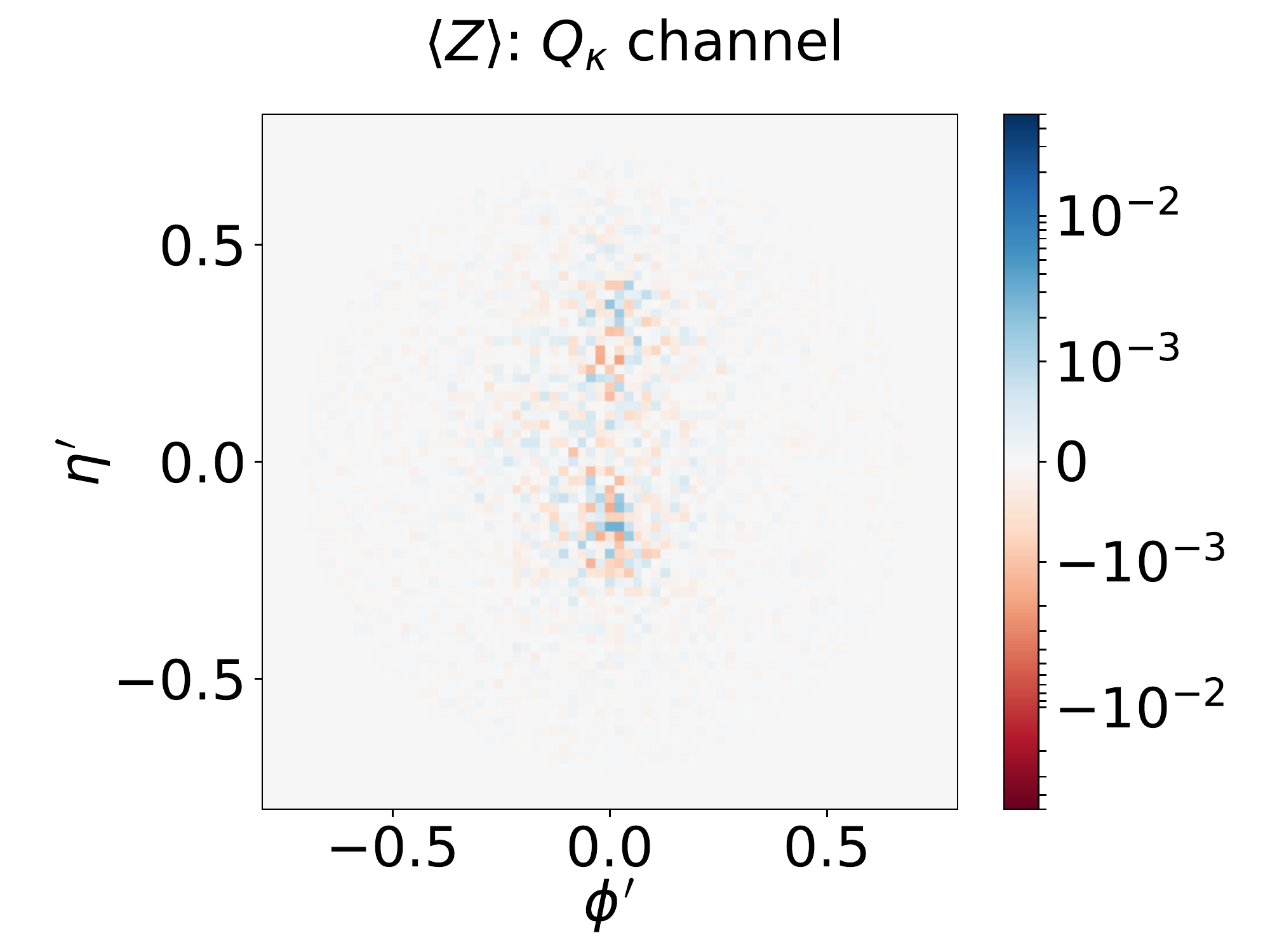}
\caption{Average of jet images in the  $\mathcal{Q}_{\kappa}$ channel, with $\kappa = 0.15$, for the three jet samples in our testing sets, after preprocessing.}
\label{fig:jetImages2}
\end{figure}

To improve the performance of our taggers, we preprocess each image, following a similar procedure as in ref.~\cite{Macaluso:2018tck}: {\it centralization}, {\it rotation} and {\it flipping}.  In figures~\ref{fig:jetImages} and \ref{fig:jetImages2}, we use $\phi'$ and $\eta'$ to denote the new coordinate system for the images after preprocessing.  Note that we do not perform {\it normalization} in the preprocessing, though it is another common preprocessing operation. Instead, we introduce a {\tt BatchNormalization} layer \cite{DBLP:journals/corr/IoffeS15} and observe a comparable (or even better) performance.

Figure~\ref{fig:jetImages} shows the average of jet images in the $p_{T}$ channel after preprocessing. For comparison, we show the $W^{+}$ average jet images in left plot and the difference between $Z$ and $W^{+}$ average jet images in the right plot. It is observed that the average $Z$ jet image is wider in the $\eta'-\phi'$ plane than the average $W$ jet image, as expected from their difference in the invariant mass.

Figure~\ref{fig:jetImages2} shows the average of jet images in the $\mathcal{Q}_{\kappa}$ channel for a fixed $\kappa$ value of $0.15$ and after preprocessing.  On average, the jet charge images are consistent with what we expect from the $W^+$, $W^-$ and $Z$-boson charges. In particular, the average $Z$-boson jet charge image is very close to zero, as the charges of the constituents tend to cancel out on the average.

\section{Methods}
\label{sec:jetChargeTagger}

In the following, we will investigate possible classification tasks, including an overall ternary problem and binary ones ($W^-$ vs.\ $W^+$ and $Z$ vs.\ $W^+$). Throughout this paper, unless otherwise specified, we will treat $W^+$ as the ``background'' in all binary classification tasks.

Here we describe the various taggers used in our work: baseline cut-based and BDT taggers that take high-level inputs ($\mathcal{M}$, $\mathcal{Q}_{\kappa}$), as well as taggers based on CNNs, which are trained on jet images formed by lower level inputs. 

\subsection{Cut-based and BDT taggers}

We first construct a cut-based tagger. We will identify the optimal values of $\kappa$ for the different binary discrimination tasks, $W^-$ vs.\ $W^+$ and $Z$ vs.\ $W^+$. For the latter, we will explore all possible simple 2D rectangular cuts in the jet charge and jet mass plane for the optimal cut-based discriminator.

We will also study a ``single-$\kappa$ BDT.'' It is built out of $\mathcal{Q}_{\kappa}$ with $\kappa$ held fixed together with the jet mass $\mathcal{M}$ . Both observables are fed into a gradient BDT implemented with the {\textsc{sklearn}}~\cite{scikit-learn} package and assuming the default parameters. We can also combine models of different $\kappa$ values to form another BDT tagger, dubbed ``multi-$\kappa$ BDT,'' similar to the ``multi-$\kappa$" jet tagger constructed in ref.~\cite{Fraser:2018ieu}. For this multi-$\kappa$ BDT, $\mathcal{M}$, $\mathcal{Q}_{\kappa}$ and $\kappa=0.2, 0.3, 0.4$ are specified as inputs.  We find that the single-$\kappa$ BDT, when taking the optimal $\kappa$ value, generally has a comparable performance as the multi-$\kappa$ BDT.  In the following sections, the single- and multi-$\kappa$ BDTs are shown as benchmark models to compare with our deep neural networks. 

\begin{figure}[ht]
\centering
\includegraphics[width=0.45\textwidth,angle=0]{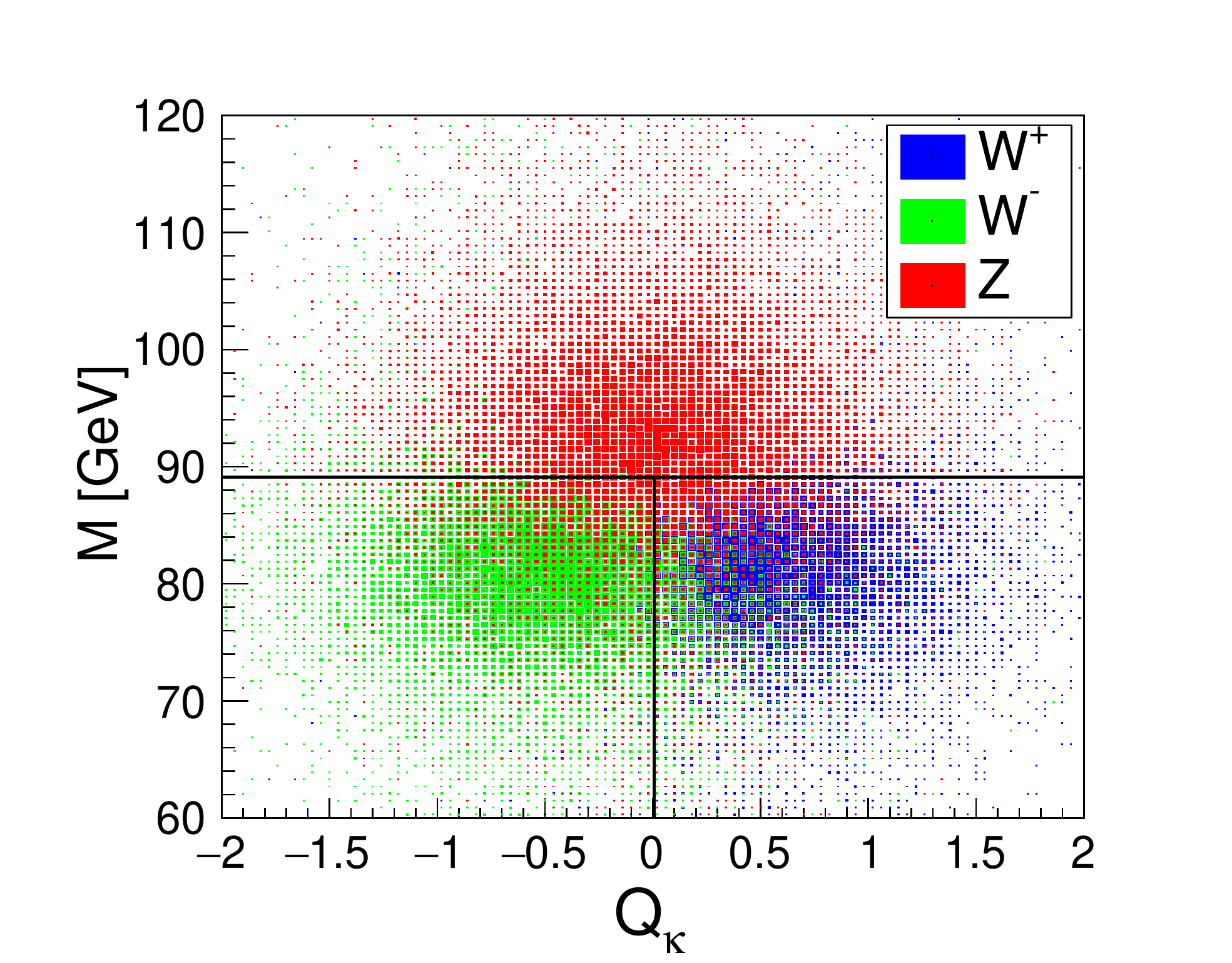}
\includegraphics[width=0.45\textwidth,angle=0]{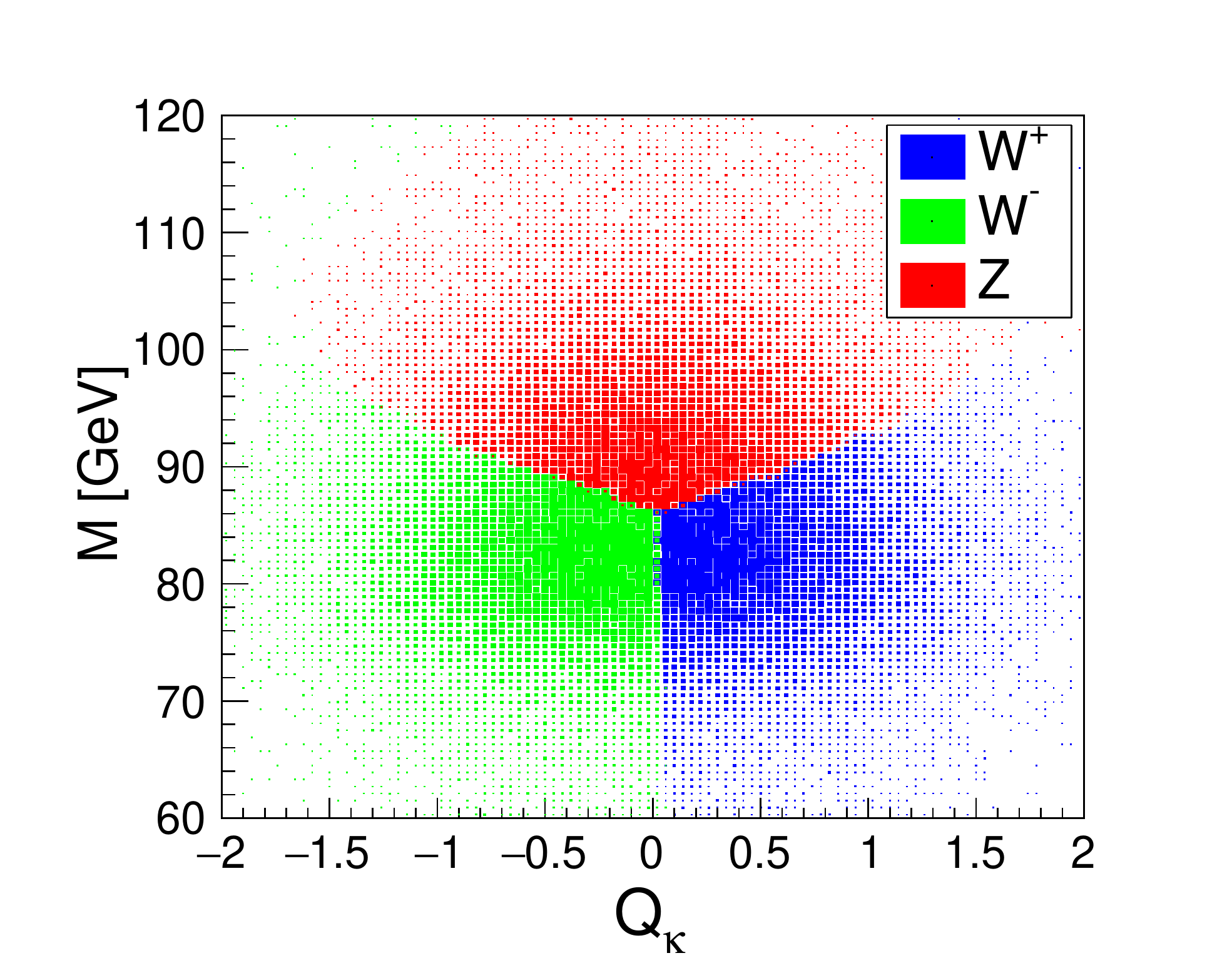}
\caption{The left plot shows the true distributions for $W^+/W^-/Z$ in the ($\mathcal{Q}_{\kappa},{\mathcal{M}}$) plane (for $\kappa=0.3$). The T-shaped lines in the left plot mark the decision boundaries of the cut-based tagger. The right plot shows the output prediction of the ternary BDT classifier in the ($\mathcal{Q}_{\kappa},{\mathcal{M}}$) plane, whose Y-shaped color boundaries match our intuition for the optimal border. 
}
\label{fig:ternaryBDT}
\end{figure}

We use these BDT taggers in both binary and ternary classifications. The prediction of the ternary single-$\kappa$ BDT classifier on the testing set can be visualized in figure~\ref{fig:ternaryBDT}. The three blobs (red, green and blue) correspond to the three classes of jets ($Z$, $W^{-}$, $W^{+}$). The single-$\kappa$ BDT nicely separates and distributes the jets as intuitively expected.  That is, to mark out the border among the three classes, the best choice would be the Y-shaped cut in the two dimensional plane.

\subsection{CNN-based taggers}

We now describe the architectures of two deep CNN models based on jet images developed in this study. We feed the two-color jet images $(p_T,Q_\kappa)$, as described in Sec.~\ref{sec:JetImages}, into the CNNs. We use the {\textsc{Keras}} library with {\textsc{TensorFlow}} backend for the implementation of the networks.

\begin{figure}[h]
\centering
\includegraphics[width=1.0\textwidth,angle=0]{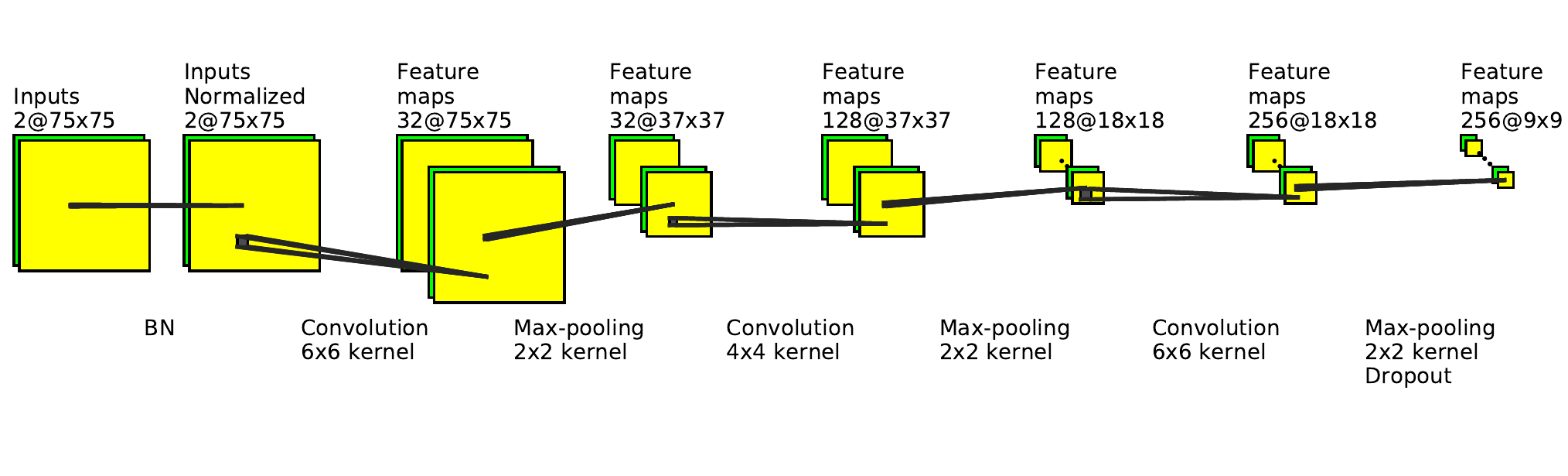}
\includegraphics[width=0.35\textwidth,angle=0]{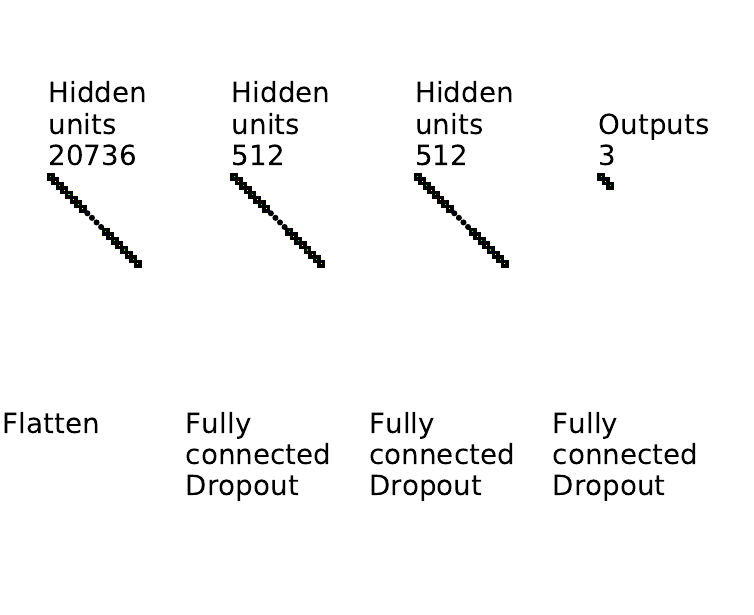}
\caption{CNN model architecture.}
\label{fig:ModelArch}
\end{figure}

Throughout the paper, the ``CNN'' label describes a network composed of 3 {\it{convolutional layers}} followed by 2 {\it{fully connected layers}}. The {\tt padding} option
is activated to enable the network to go deeper. To prevent overfitting, {\tt regularizers} are used as well as {\it{dropout layers}}. For training, we use {\tt Adam}~\cite{Kingma:2014vow} as the optimizer algorithm.\footnote{We also tried the {\tt SGD} optimizer and the performance is roughly the same as {\tt Adam}.} 
In figure~\ref{fig:ModelArch}, we show the architecture of the CNN, with detailed model configuration parameters summarized in table~\ref{tab:Networks}. 

The second deep neural network we have considered is a composite design consisting of two CNNs with asymmetrical depths, which we call ``CNN$^2$.'' The machine learning in the $p_{T}$ and $\mathcal{Q}_{\kappa}$ channels is done in parallel, {\it i.e.}, we feed the $p_{T}$ images and the $\mathcal{Q}_{\kappa}$ images into two separate CNNs that differ in structure. These are combined at the end in a merge layer.

The CNN$^2$ architecture is motivated by our finding that the depth of the CNN network is limited by the classification task between $W^+$ and $W^-$.  If the network is made too deep, the $W^+/W^-$ tagger tends to overfit. This makes sense, since  the $W^+$ and $W^-$ jet images are identical except in the sign of the $\mathcal{Q}_{\kappa}$ channel.  Therefore, it requires fewer convolutional layers to capture the difference between the two. On the other hand, a deeper network structure does help a lot in successfully identifying the $Z$ boson as the signal. The $Z$ samples differ from the other two in the spatial distribution (substructure) of the constituents. If we enhance the resolution/ability of the CNN's pattern recognition, it is natural that the CNN could do better in the $Z$ discrimination. Therefore, there seems to be a trade-off in the ternary classification problem. The CNN$^2$ architecture is an attempt to have our cake and eat it too. 

\begin{figure}[ph]
\centering
\includegraphics[width=0.65\textwidth,angle=0]{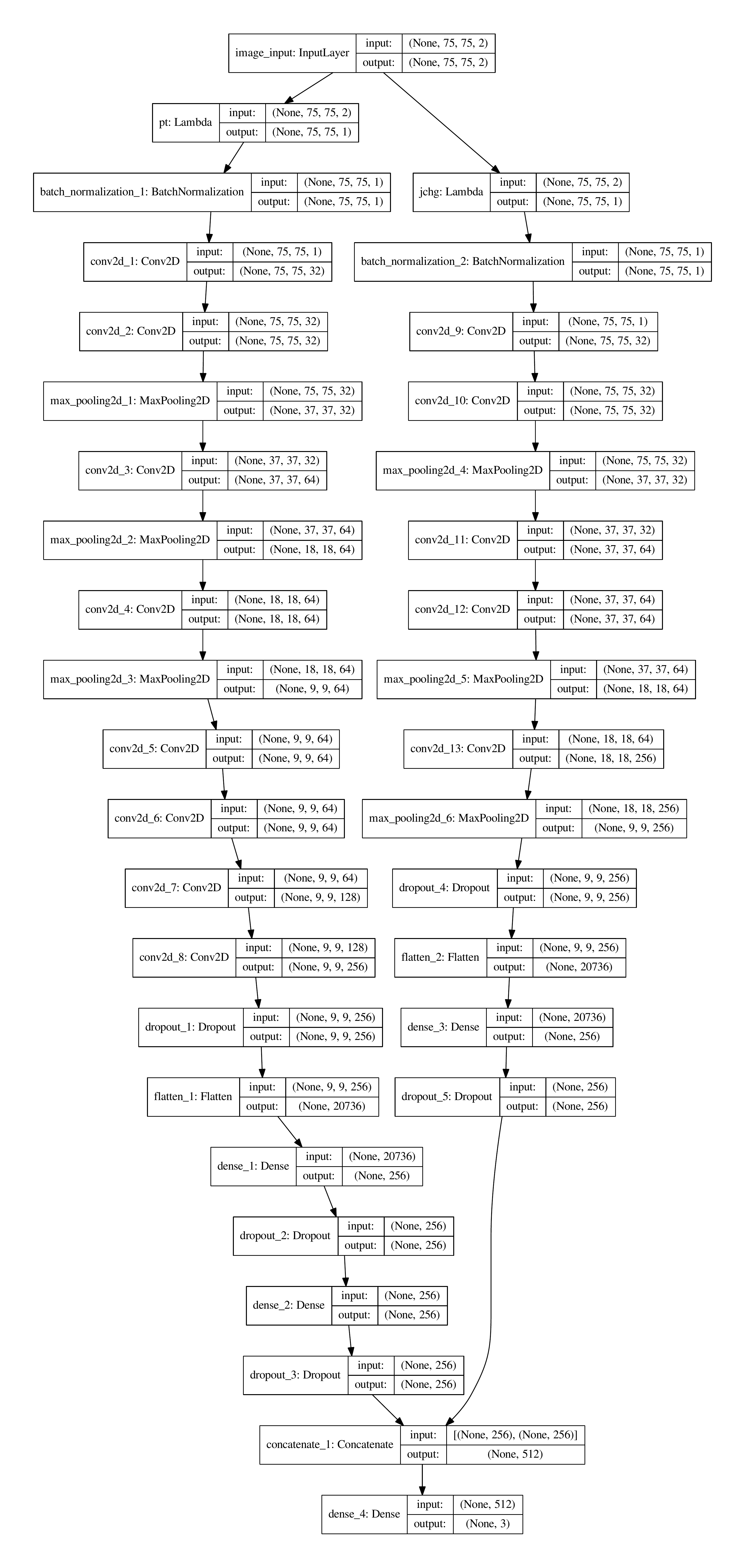}
\caption{Detailed model architecture of CNN$^{2}$.}
\label{fig:CNN2model}
\end{figure}

After investigations on the model structure and seeing performance trends in the different classification problems, we have chosen the CNN$^2$ architecture detailed in figure~\ref{fig:CNN2model} and table~\ref{tab:Networks}. The one dealing with the $\mathcal{Q}_{\kappa}$ images is shallower. Based on the observed performance, we keep on stacking up to 8 convolutional layers for the other CNN that processes the $p_{T}$ images.

\begin{table}[h]
\footnotesize
\centering
\begin{threeparttable}
\begin{tabular}{c|c|c|c}
\hline\hline
& CNN & \multicolumn{2}{c}{CNN$^2$}
\\
\hline
Image & \multicolumn{3}{c}{($75\times75$) pixels within $(|\eta|\leq 0.8$, $|\phi| \leq 0.8)$}
\\
\hline
Channels & $p_{T}$, $\mathcal{Q}_{\kappa}$ & $p_{T}$ & $\mathcal{Q}_{\kappa}$
\\
Architecture & BN-32C6-MP2-128C4- & BN-32C3-32C3-MP2- & BN-32C3-32C3-MP2-
\\
& MP2-256C6-MP2-512N- & 64C3-MP2-64C3-MP2- & 64C4-64C4-MP2-256C6-
\\
& 512N & 64C3-64C3-128C5-256C5- & MP2-256N
\\
& & 256N-256N &
\\
\hline
Settings & \multicolumn{3}{c}{{\texttt{Relu}} Activation, \texttt{Padding=same}, {\it{Dropout}} = 0.5, l2 \texttt{Regularizer} = 0.01}
\\
Preprocessing & \multicolumn{3}{c}{\it{Centralization, Rotation, Flipping}}
\\
Training & \multicolumn{3}{c}{Adam Optimizer, \texttt{Minibatchsize}=512, {\it{Cross entropy}} loss}
\\
\hline\hline
\end{tabular} 
\caption{Summary of the configurations of our CNN taggers.}
\label{tab:Networks}
\end{threeparttable}
\end{table}

\section{$W^{-} / W^{+}$ binary classification}
\label{sec:binary_wpwn}

We begin with a study of the $W^-/W^+$ classification. Since the signal and background differ only in their charges, the only useful quantity here is the jet charge $Q_\kappa$. 
An important aspect of the jet charge variable defined in eq.~\eqref{eq:jetcharge} is that it depends on a parameter $\kappa$ which specifies how the contributing charges are $p_T$-weighted. Nothing {\it a priori} tells us which value of $\kappa$ to use, and different tasks may prefer different values of $\kappa$. In the following subsection we examine this issue for our various taggers.

\subsection{Determining $\kappa$}
\label{sec:KappaStudy}

\begin{figure}[ht]
\centering
\includegraphics[width=\textwidth,angle=0]{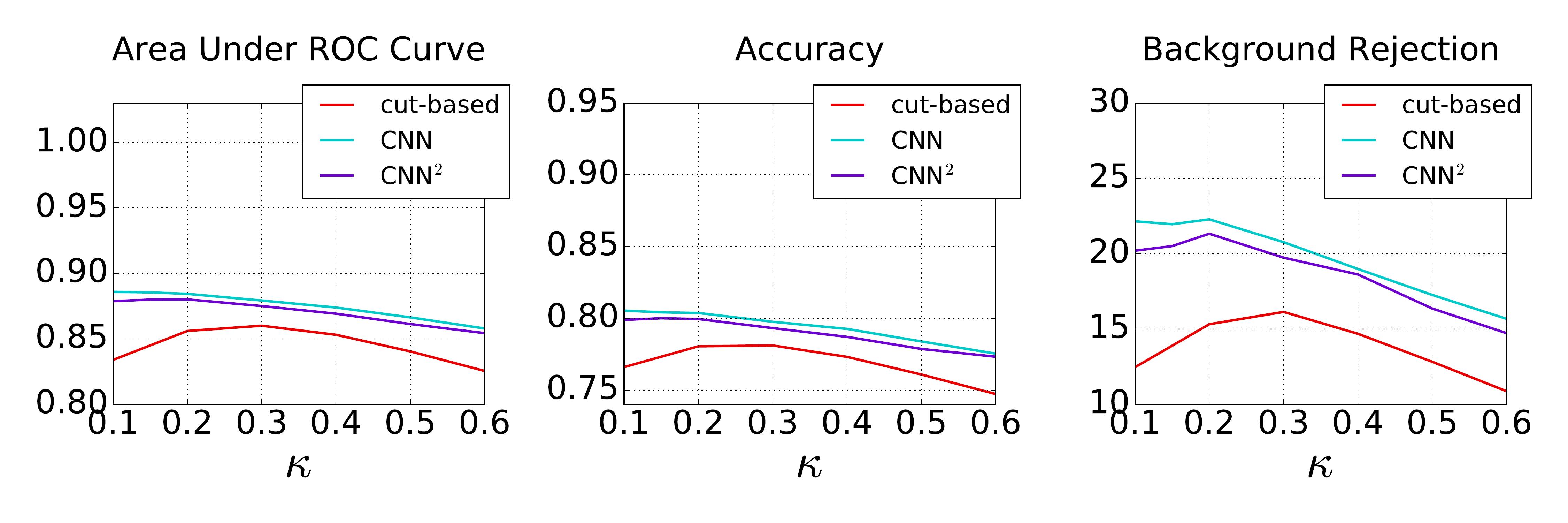}
\caption{Summary of the performance as a function of $\kappa$ in the range $[0.1,0.6]$ for a binary classification task to discriminate $W^{-}$ from $W^{+}$ for all taggers. The three metrics are the AUC (left), accuracy (middle) and background rejection (right).}
\label{fig:jetChargeSummary2}
\end{figure}

In figure~\ref{fig:jetChargeSummary2}, we show the trend in performance when varying $\kappa$ for the $W^{-}$ vs.\ $W^{+}$ classification. 
We show three typical metrics for evaluating the algorithm's performance: area under the curve (AUC), best accuracy (ACC) and background rejection at a $50 \%$ signal efficiency working point ($1/\epsilon_{b}|_{\epsilon_{s} = 50 \%}$, denoted by R50). Deep learning taggers as well as the cut-based reference tagger are plotted together for comparison. Note in particular that the performance of single-$\kappa$ BDT, not shown in these plots, would trivially reduce to that of the cut-based tagger in the $W^{-}/W^{+}$ binary classification task, in which case the $\mathcal{M}$ information does not provide additional discriminative power.

We find in figure~\ref{fig:jetChargeSummary2} that the performance does depend on the choice of $\kappa$. It is also interesting to see that the deep learning taggers have a qualitatively different $\kappa$ dependence from the traditional (cut-based) taggers, while little difference exists between the two CNN models.  The former are always better than the latter and have a smaller optimal $\kappa$.

Based on these results, we determine the optimal $\kappa$ in each of the tagger definitions in the following sections. A value of $\kappa=0.3$ is fixed for the single-$\kappa$ BDT reference tagger, and $\kappa=0.15$ for the CNN taggers.

\subsection{Comparison of taggers}

\begin{figure}[h]
\centering
\includegraphics[width=0.45\textwidth,angle=0]{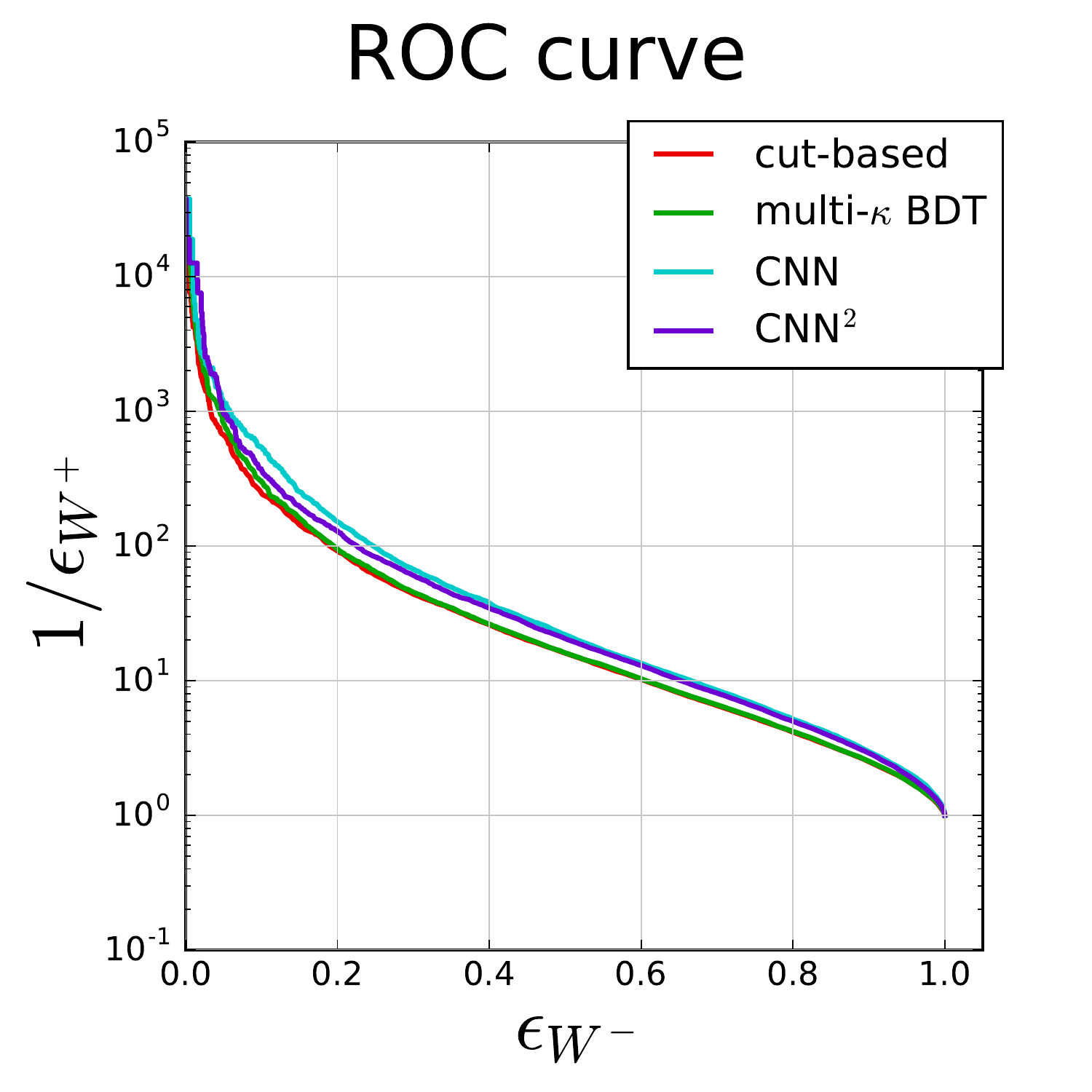} 
\includegraphics[width=0.45\textwidth,angle=0]{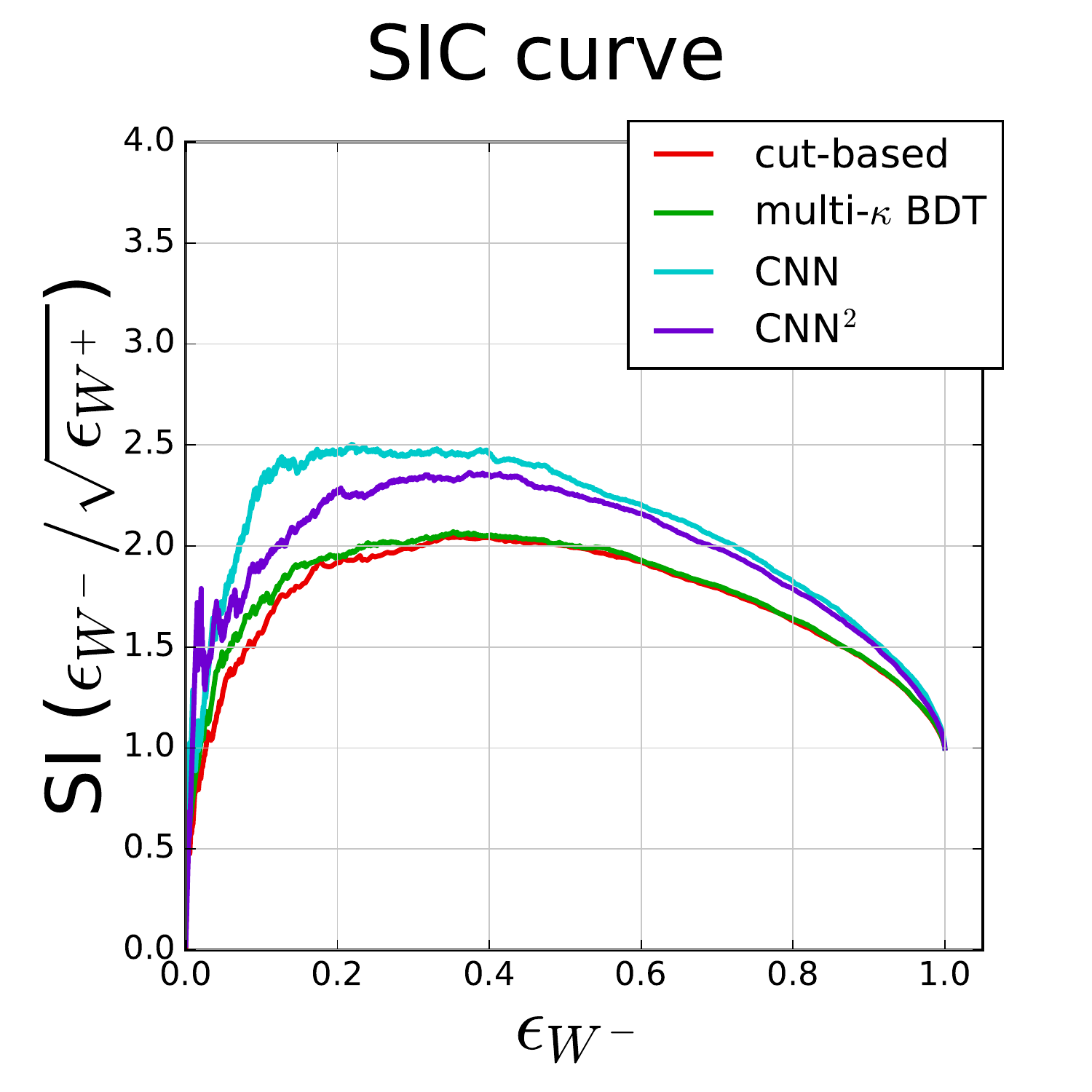} 
\caption{ROC (left) and SIC (right) curves for the binary classification to discriminate $W^{-}$ from $W^{+}$ for all taggers, except for the single-$\kappa$ BDT.}
\label{fig:800-SIC-Binary_wpwnFinal}
\end{figure}

Having fixed the value of $\kappa$ in each tagger, we are now ready to compare the various tagging methods. Table~\ref{tab:CNNw_binary} shows the performance metrics described above (AUC, ACC and R50). In figure~\ref{fig:800-SIC-Binary_wpwnFinal}, the left plot shows the Receiver Operating Characteristic (ROC) curves and the right plot shows the Significance Improvement Characteristic (SIC) curves. While all the ROC curves seem to be close to one another, one can readily see a clear benefit from employing deep learning in the SIC curves. The improvement in background rejection rate is around 30-40\% across a wide range of signal efficiencies. It is also worth noting that for the $W^{-}/W^{+}$ task, there is no particular gain in using the CNN$^{2}$ structure; in fact, the performance is slightly worse. 

\begin{table}[h]
\begin{center}
\begin{tabular*}{0.55\textwidth}{c|ccc}
\bottomrule
           &      \hspace{0.2cm}   R50 &\hspace{0.25cm}   AUC\hspace{0.25cm}  &\hspace{0.25cm}   ACC \hspace{0.25cm} \\
           \hline
cut-based 		   &  16.1372  &  0.8600  &  0.7811  \\
multi-$\kappa$ BDT         &  16.0960  &  0.8615  &  0.7820  \\
CNN       		   &  21.9559  &  0.8855  &  0.8042  \\
CNN$ ^2$  		   &  20.5057  &  0.8800  &  0.8000  \\
\toprule 
\end{tabular*} 
\caption{Performance metrics for all taggers, except for the single-$\kappa$ BDT, in a $W^-/W^+$ binary classification task.}
\label{tab:CNNw_binary}
\end{center}
\end{table}

Although differing in the details, it is still instructive to compare with the previous work on jet charge and deep learning~\cite{Fraser:2018ieu}, which focuses on up/down quark jet discrimination. Our performance gain from BDT to deep learning turns out to be quite comparable.
For the $1000$-GeV benchmark scenario, their CNN's background rejection rate at $50 \%$ signal efficiency grows by about $40 \%$ relative to their ``$\kappa$ and $\lambda$ BDT'' reference tagger, as given in table 1 of ref.~\cite{Fraser:2018ieu}. 
Our result shows the same amount of enhancement, as seen in table~\ref{tab:CNNw_binary}.

\section{$Z / W^{+}$ binary classification}
\label{sec:binary_wz}

Next we turn to the $Z$ vs.\ $W^+$ binary classification task. ($Z$ vs.~$W^-$ would obviously have the same result because of charge symmetry.) 
Here we are primarily interested in how much the jet charge observable adds to the discriminative power, compared to just the information computable from the four vectors ({\it e.g.}, the jet mass).

\subsection{Determining $\kappa$}

\begin{figure}[h]
\centering
\includegraphics[width=\textwidth,angle=0]{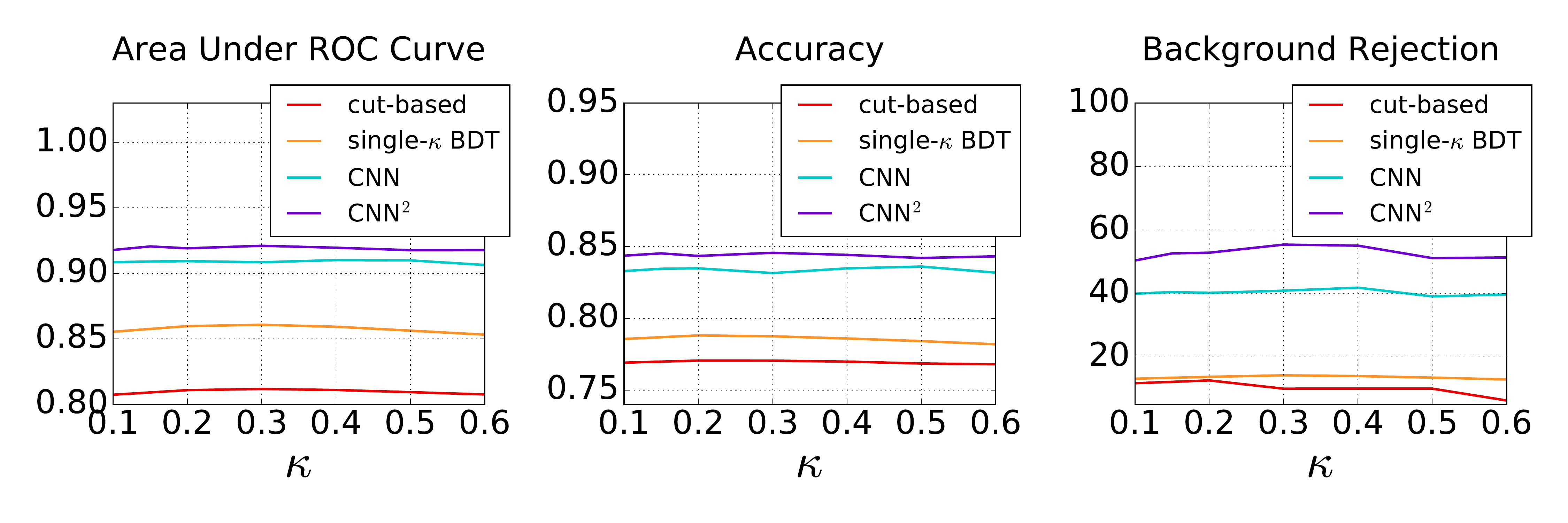}
\caption{Same as figure~\ref{fig:jetChargeSummary2}, but for a binary classification task to discriminate $Z$ from $W^{+}$ for all taggers.}
\label{fig:jetChargeSummaryWZ}
\end{figure}

The performance of various taggers as a function of $\kappa$ is shown in figure~\ref{fig:jetChargeSummaryWZ}. Unlike the $W^-/W^+$ classification, the dependence in $\kappa$ is very mild here. We will keep using the same values of optimal $\kappa$ as before for simplicity.

\subsection{Comparison of taggers}

\begin{figure}[h]
\centering
\includegraphics[width=0.45\textwidth,angle=0]{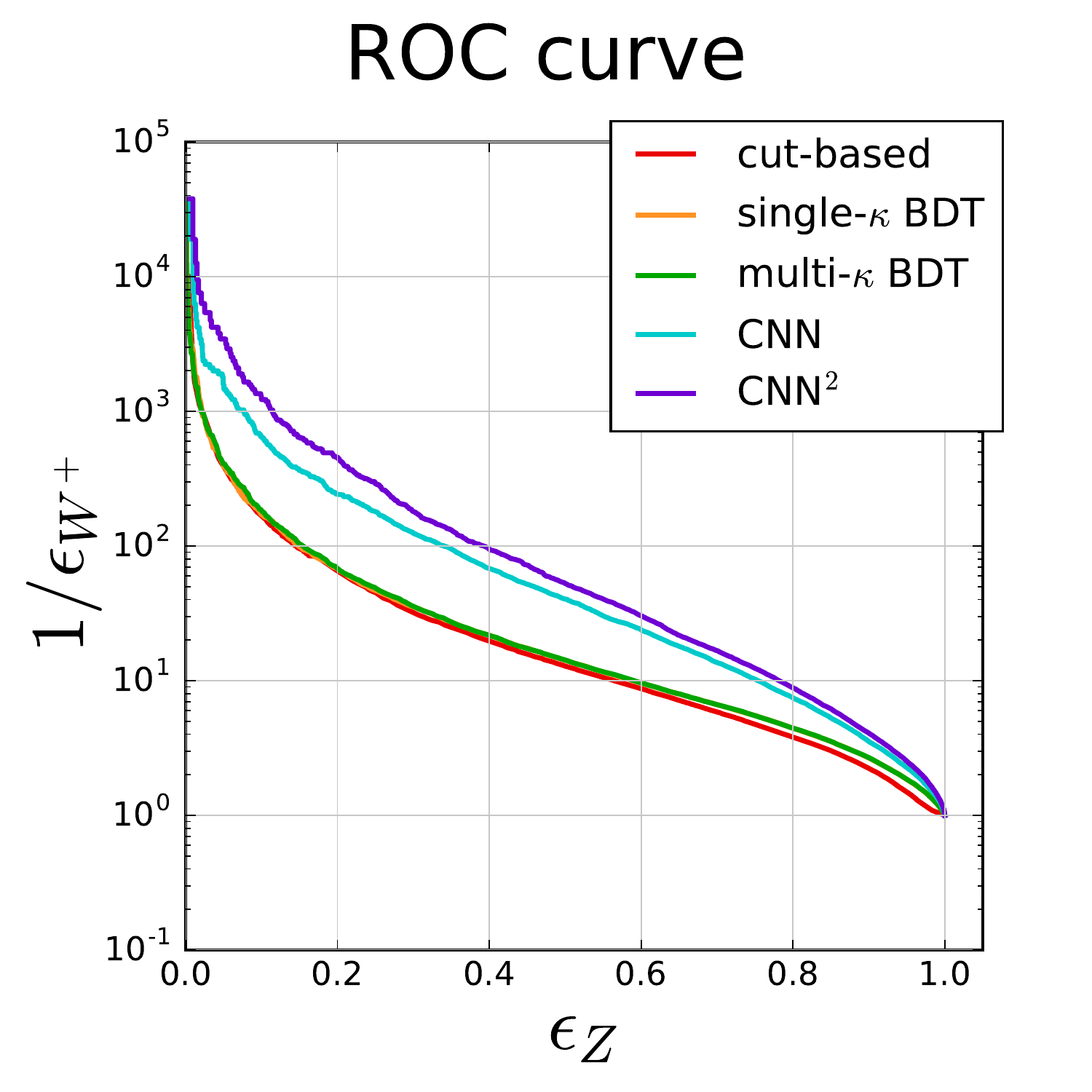}
\includegraphics[width=0.45\textwidth,angle=0]{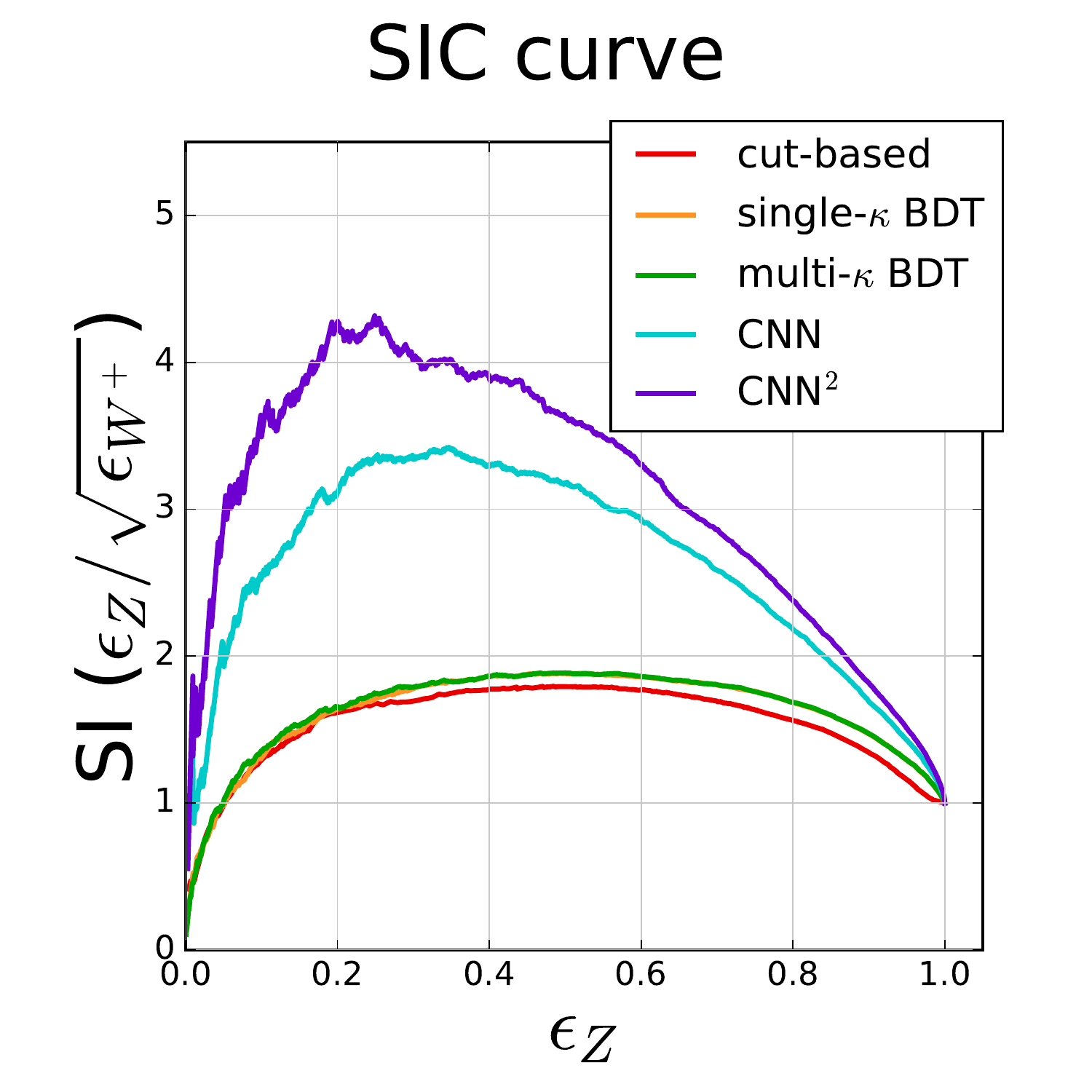}
\caption{ROC (left) and SIC (right) curves for a binary classification discriminating $Z$ from $W^{+}$ for all taggers.}
\label{fig:800-SIC-BinaryWZFinal}
\end{figure}

Figure~\ref{fig:800-SIC-BinaryWZFinal} shows the ROC curves (left plot) and the SIC curves (right plot) of the different taggers in the binary task of distinguishing $Z$ from $W^{+}$. Table~\ref{tab:CNNz_binary} gives the three performance metrics.

Compared to the $W^-/W^+$ classification task, the benefit from deep learning in the current case is much greater, with an improved background rejection rate at 50\% signal efficiency of as much as $\sim 2.85$. This is perhaps unsurprising, since our cut-based and BDT methods do not include any jet substructure variables. As noted before in figure~\ref{fig:jetImages}, in the $W^{-} / W^{+}$ classification task, the samples have identical average distribution in the $p_T$ images and differ only in pixel intensity in the $\mathcal{Q}_{\kappa}$ channel.  However, the $Z$ and $W^+$ events have distinct spatial distributions in the jet's constituents, in addition to the charge difference. Since the CNN naturally learns spatial differences, it will naturally show a big improvement over methods that do not include any substructure information.

We also note that for the $Z/W^{+}$ task, the CNN$^{2}$ tagger with the architecture tailored for better $Z$ identification outperforms the CNN tagger by a sizable amount.

\begin{table}[h]
\begin{center}
\begin{tabular*}{0.55\textwidth}{c|ccc}
\bottomrule
           &      \hspace{0.2cm}   R50 &\hspace{0.25cm}   AUC\hspace{0.25cm}  &\hspace{0.25cm}   ACC \hspace{0.25cm} 	\\
           \hline
cut-based 		   &  9.9590   &  0.8118  &  0.7705  \\
single-$\kappa$ BDT        &  14.1638  &  0.8608  &  0.7875  \\
multi-$\kappa$ BDT         &  14.2383  &  0.8611  &  0.7880  \\       
CNN          		   &  40.4205  &  0.9091  &  0.8345  \\
CNN$ ^2$     		   &  52.6028  &  0.9206  &  0.8452 \\
\toprule 
\end{tabular*} 
\caption{Performance metrics for all taggers in a binary $Z/W^+$ classification task.}
\label{tab:CNNz_binary}
\end{center}
\end{table}

\begin{figure}[h]
\centering
\includegraphics[width=0.45\textwidth,angle=0]{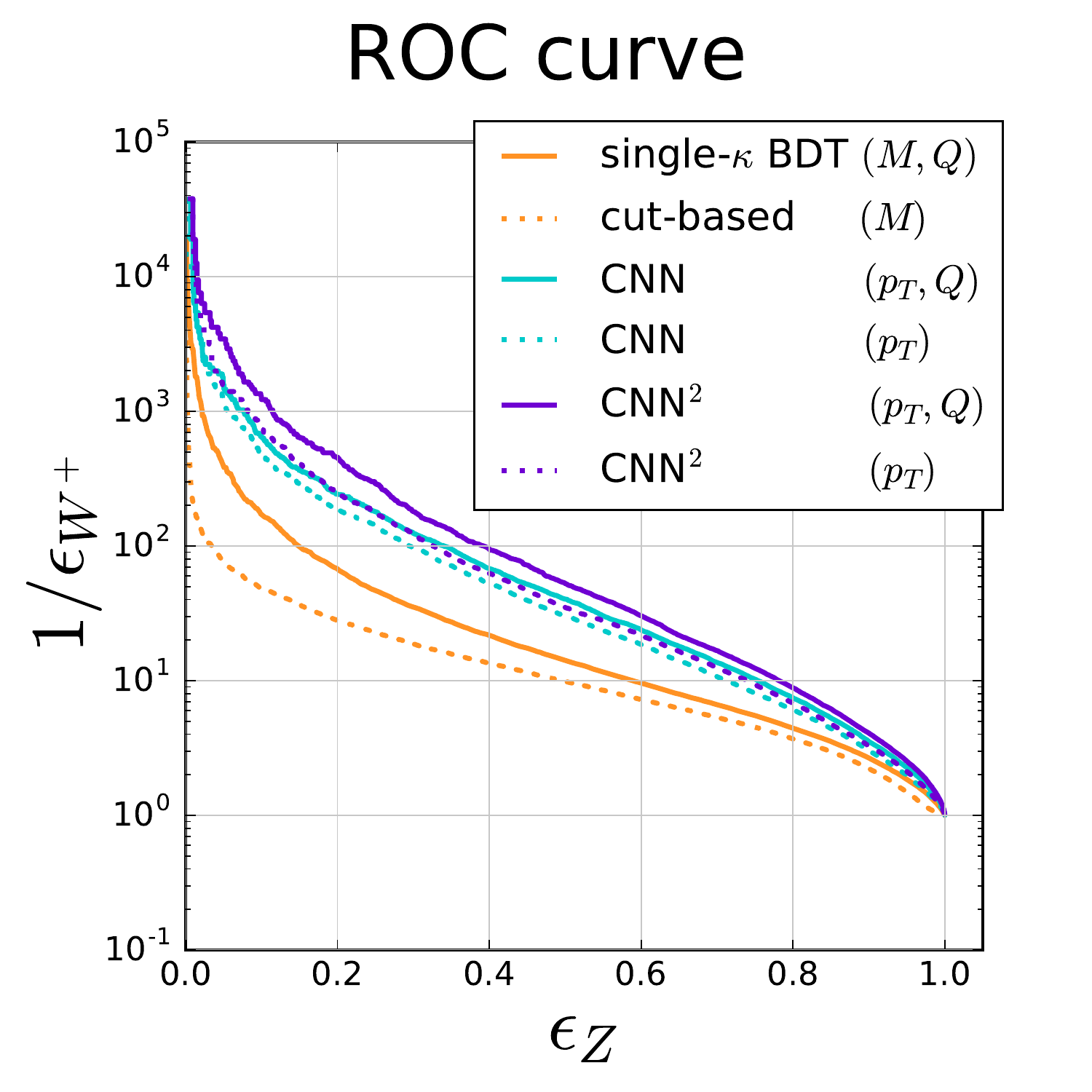}
\includegraphics[width=0.45\textwidth,angle=0]{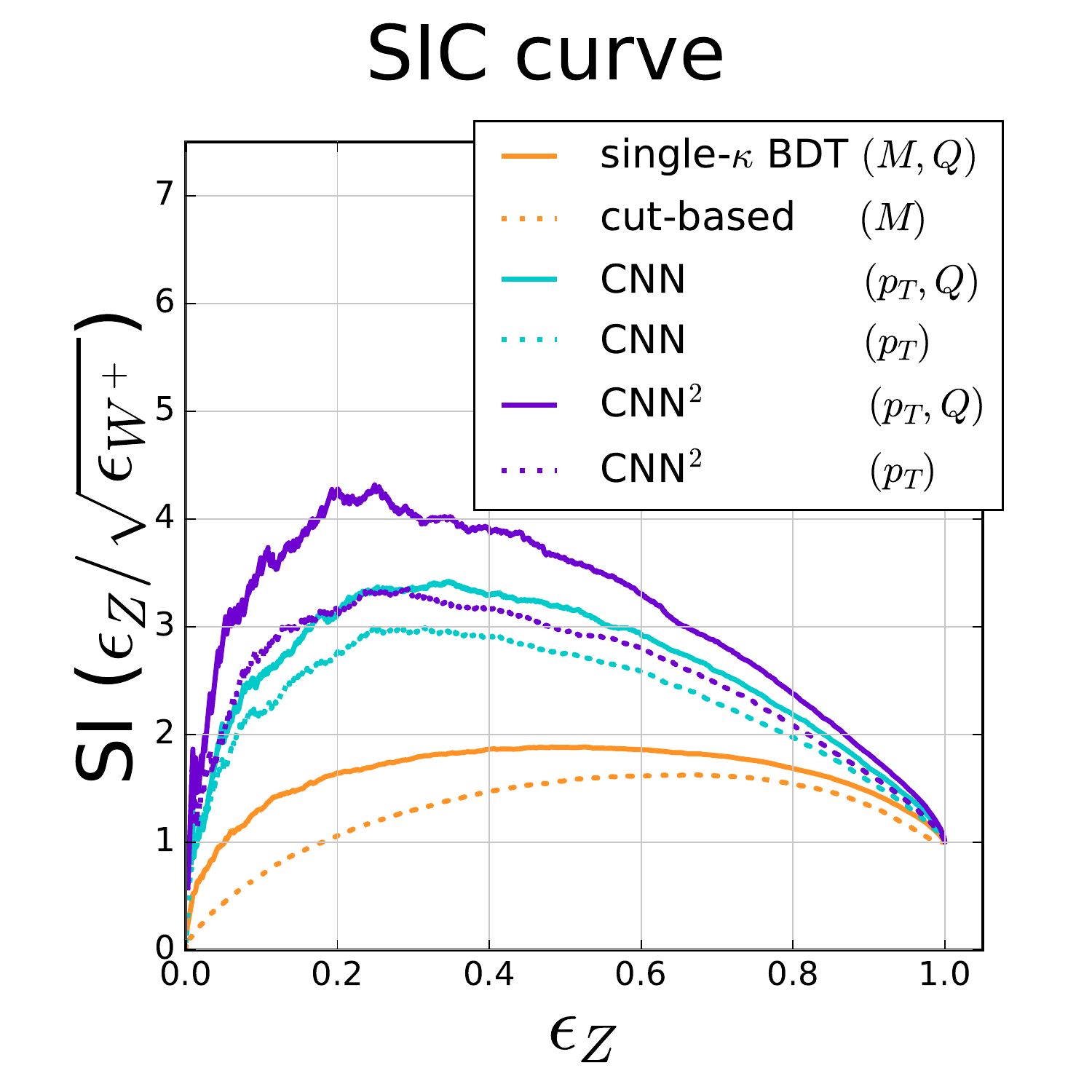}
\caption{ROC (left) and SIC (right) curves for $Z/W^+$ binary classification using taggers with different numbers of input channels. The dotted lines are for one channel only, either $\mathcal{M}$ (for the reference cut-based and BDT taggers) or $p_{T}$ (for the CNN taggers), and the solid lines are for those with two channels ($\mathcal{M}+\mathcal{Q}_{\kappa}$ or $p_{T}+\mathcal{Q}_{\kappa}$) and correspond to the results in the previous section. The single-$\kappa$ BDT tagger is compared with the cut-based tagger because the BDT with only one input channel reduces to the cut-based tagger.}
\label{fig:SICZsplit}
\end{figure}

We now scrutinize the role of jet charge in $Z$ vs.\ $W^+$ discrimination. 
Figure~\ref{fig:SICZsplit} shows the performance of taggers with either one or two input channels.  
We split our predictions from $\mathcal{M}$ to $(\mathcal{M},\mathcal{Q_{\kappa}})$ in the BDT case, or from $p_{T}$ to $(p_{T},\mathcal{Q}_{\kappa})$ for the CNNs. 
Comparing the solid curves to the dotted ones, we see that all the three taggers get significant improvements after adding the $\mathcal{Q}_{\kappa}$ information.

The role of jet charge in boosted $Z$ vs.\ $W^+$ discrimination was previously studied by ATLAS~\cite{Aad:2015eax}. However, the ATLAS study is different from ours in details: it focuses on a different signal process, $W'\to W Z$; its jet samples are defined differently; and what they construct is a likelihood tagger with the ${\mathcal{M}}$ and $\mathcal{Q}_{\kappa}$ as the inputs. Nevertheless, it is instructive to compare our results to theirs. Throughout a wide range of working points, both the ATLAS tagger and our CNN tagger attain an additional 30\% enhancement in the background rejection rate after further incorporating $\mathcal{Q}_{\kappa}$ information. In the high signal efficiency region, our taggers seem to enjoy a larger gain by introducing the $\mathcal{Q}_{\kappa}$ channel.

\section{Ternary $W^{+} / W^{-} / Z$ classification}
\label{sec:3to2}

Finally, we turn to the ultimate task of a full classification of boosted weak gauge bosons ($W^{+}/ W^{-}/ Z$). We will quantify how well the full ternary classification performs. Then we will see how to reduce to the binary classifications described in the previous sections.  For the choice of $\kappa$ for the CNN taggers, we will continue to use the same values of $\kappa$ as before because they are also optimal for the ternary classifier.

We summarize and compare the performance of the ternary taggers according to two metrics: their overall accuracy, defined as (number of correct predictions)/(total number of instances);\footnote{For BDT- and CNN-based taggers, the  largest class probability is taken to be the prediction.} and a ``one-against-all" metric which binarizes the task, {\it i.e.}, singling out one class as the ``signal'' and treating all the others as the ``background.'' 

\begin{figure}[h]
\centering
\includegraphics[width=0.45\textwidth,angle=0]{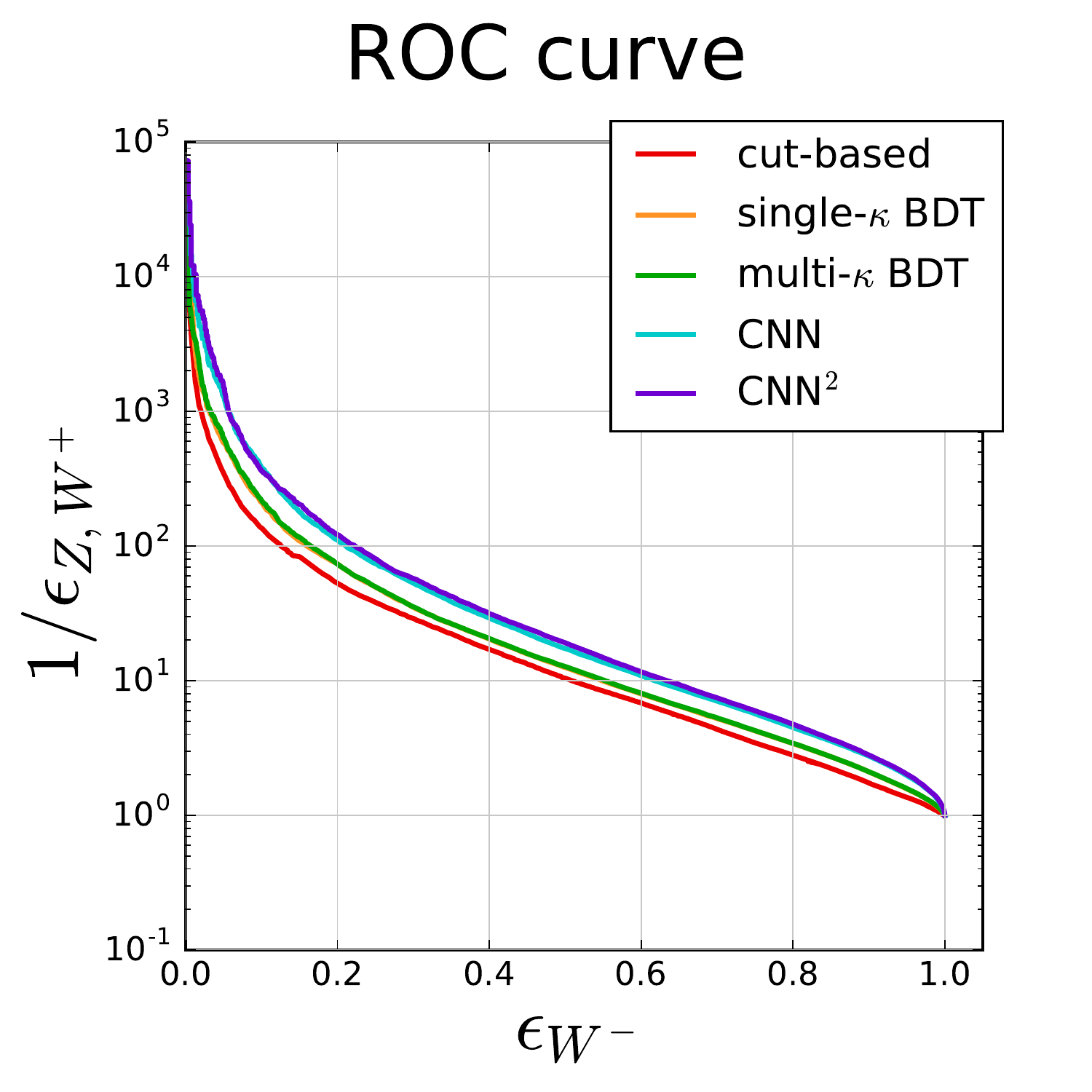}
\includegraphics[width=0.45\textwidth,angle=0]{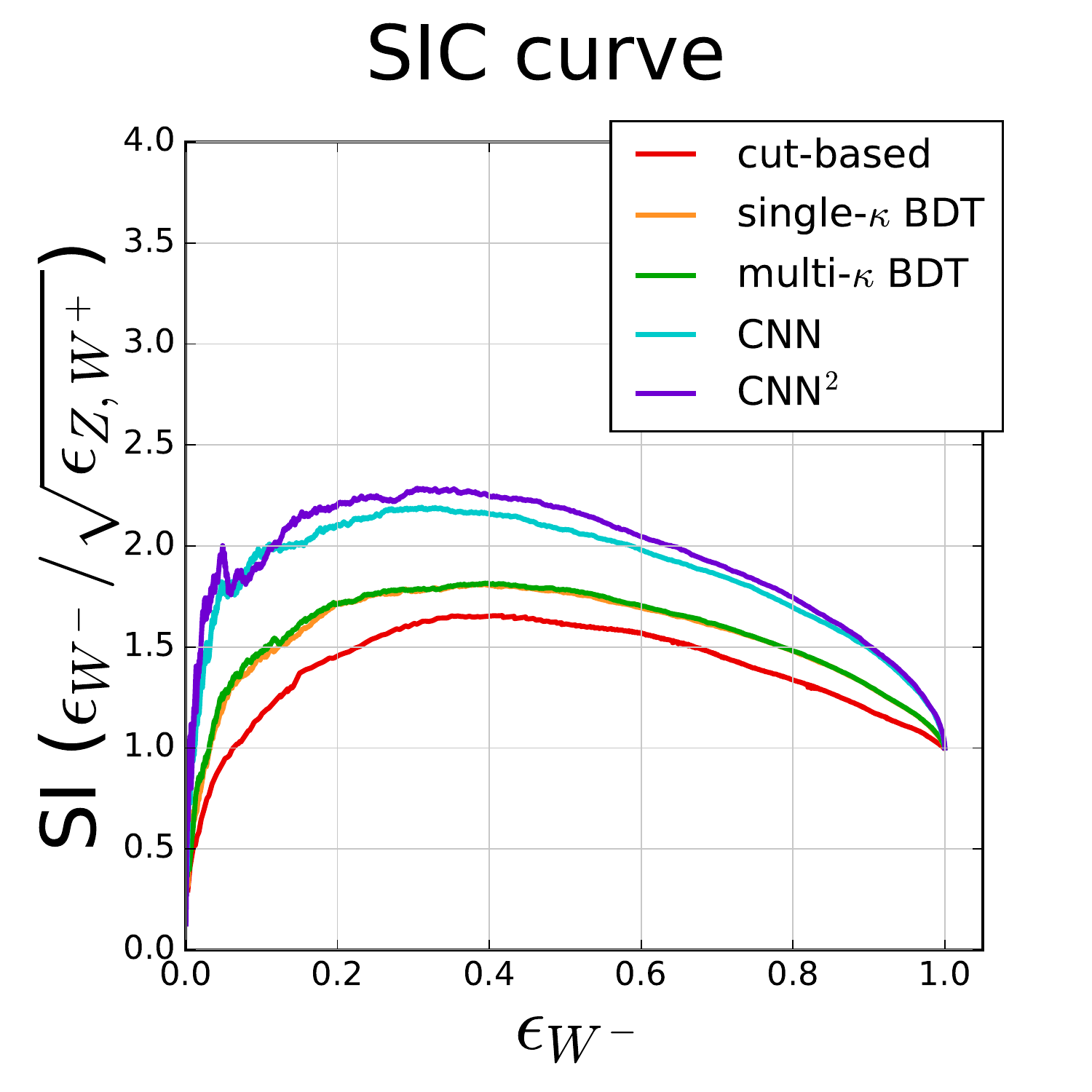}
\caption{ROC (left) and SIC (right) curves for a ternary classification discriminating $W^{-}$ from $(W^{+},Z)$ for all the taggers.}
\label{fig:800-wp-SIC-TernaryFinal}
\end{figure}

\begin{figure}[h]
\centering
\includegraphics[width=0.45\textwidth,angle=0]{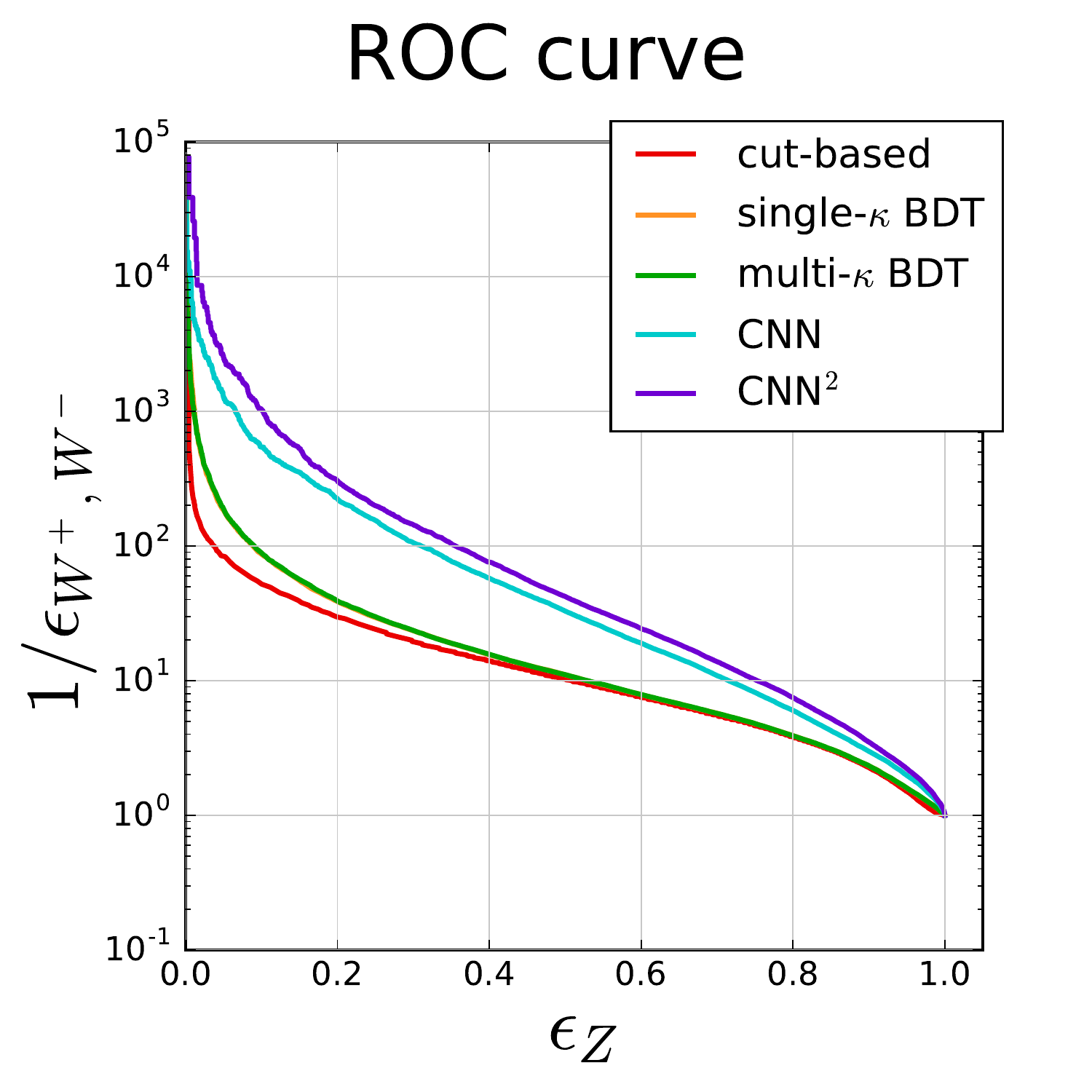}
\includegraphics[width=0.45\textwidth,angle=0]{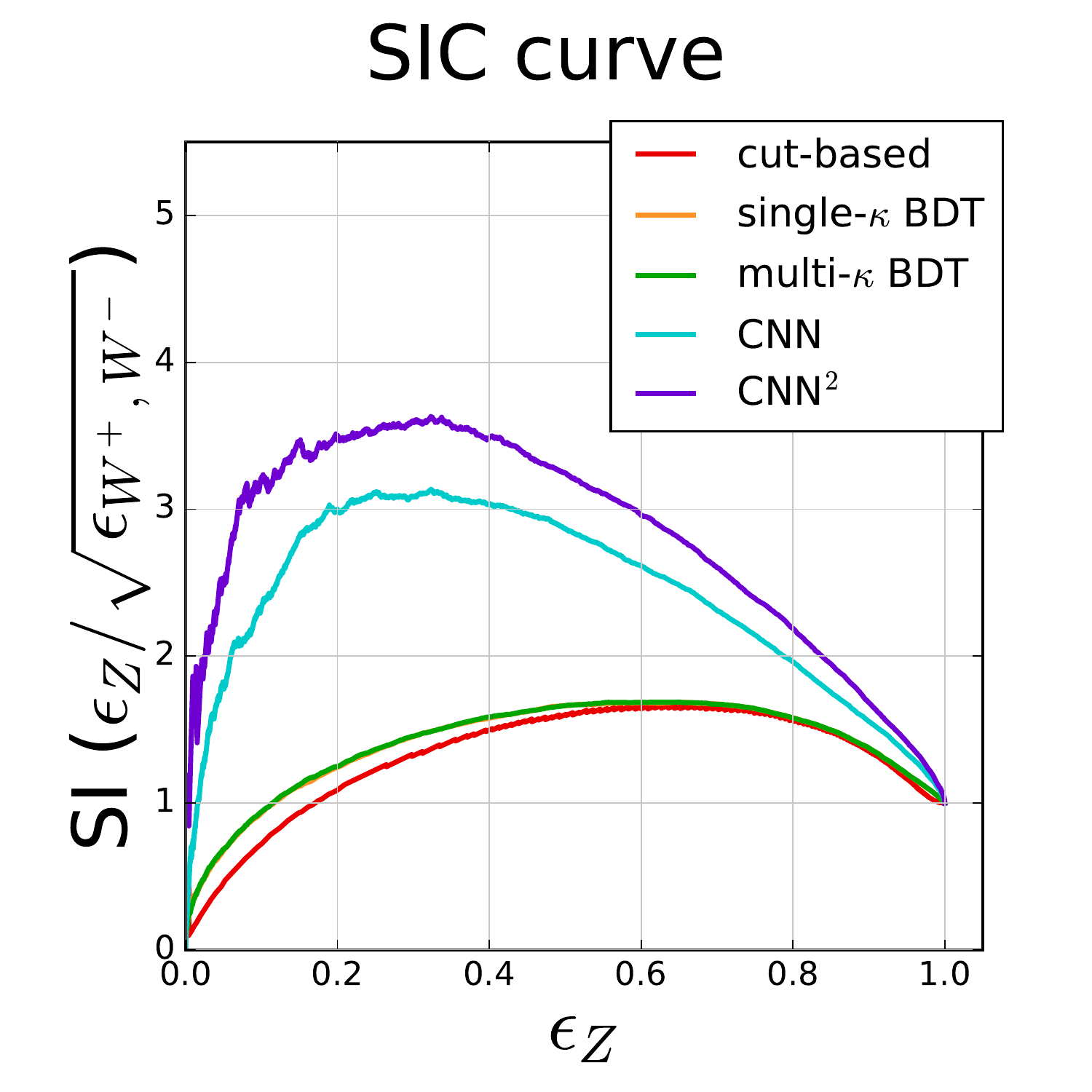}
\caption{ROC (left) and SIC (right) curves for a ternary classification discriminating $Z$ from $W$s for all the taggers. }
\label{fig:800-z-SIC-TernaryFinal}
\end{figure}

Figures~\ref{fig:800-wp-SIC-TernaryFinal} and \ref{fig:800-z-SIC-TernaryFinal} plot the ROC and SIC curves resulting from treating $W^-$ and $Z$ as the signals, respectively.\footnote{The $W^{+}$-against-all metric is omitted here, because the results and tendencies are very comparable to the $W^{-}$-against-all metric by symmetry. } The corresponding metrics are given in table~\ref{tab:800-SIC-Ternary}.
Our results show that the CNN$^{2}$ generally performs better than CNN. Thanks to the parallel structure, the network has a sufficient depth for a sizable improvement in the $Z$-signal performance, while having comparable or better performance than the CNN in the $W^-$ (or $W^+$) discrimination.
 
\begin{table}[h]
\begin{center}
\begin{tabular*}{0.93\textwidth}{c|c|ccc|ccc}
\bottomrule
  &  overall   &   \multicolumn{3}{c|}{signal: $W^{-}$} 
  &  \multicolumn{3}{c}{signal: $Z$} 
\\
           &  \hspace{0.1cm} ACC \hspace{0.1cm}
           &  \hspace{0.02cm}  R50 &\hspace{0.02cm}  AUC\hspace{0.02cm}  &\hspace{0.02cm}  ACC \hspace{0.02cm} 
           &  \hspace{0.02cm}  R50 &\hspace{0.02cm}  AUC\hspace{0.02cm}  &\hspace{0.02cm}  ACC \hspace{0.02cm} 	\\
           \hline
cut-based             & 0.6581 & 8.0262  &  0.7893  &  0.7643 & 10.0882 &  0.8233  &  0.7839 \\
single-$\kappa$ BDT   & 0.6667 & 12.5230 &  0.8339  &  0.7576 & 11.0726 &  0.8363  &  0.7725 \\
multi-$\kappa$ BDT    & 0.6675 & 12.7115 &  0.8348  &  0.7579 & 11.0678 &  0.8366  &  0.7726\\
CNN 		      & 0.7197 & 17.3403 &  0.8715  &  0.7890 & 32.8981 &  0.8936  &  0.8170\\
CNN$^2$ 	      & 0.7318 & 19.0907 &  0.8764  &  0.7950 & 42.1927 &  0.9088  &  0.8334\\
\toprule 
\end{tabular*} 
\caption{Performance metrics for all taggers in the ternary classification task.}
\label{tab:800-SIC-Ternary}
\end{center}
\end{table}

\subsection{Comparison with binary taggers}
\label{sec:3to2CmpBinary} 

We expect that a multi-class classification task should be able to fully recover the binary classification performance after an appropriate projection. If the multi-class NN output is supposed to approximate the class probability $P_i(x)$, where $x$ is a data point and $i=1,\dots,N$ is the class label, then the projection to binary classification between class $i$ and class $j$ is simply:
\begin{equation}
P_i^{i\, \, {\rm or}\,\, j}(x) = \frac{P_i(x)}{P_i(x)+P_j(x)} ~.
\label{eq:probability}
\end{equation}

\begin{figure}[h]
\centering
\includegraphics[width=0.45\textwidth,angle=0]{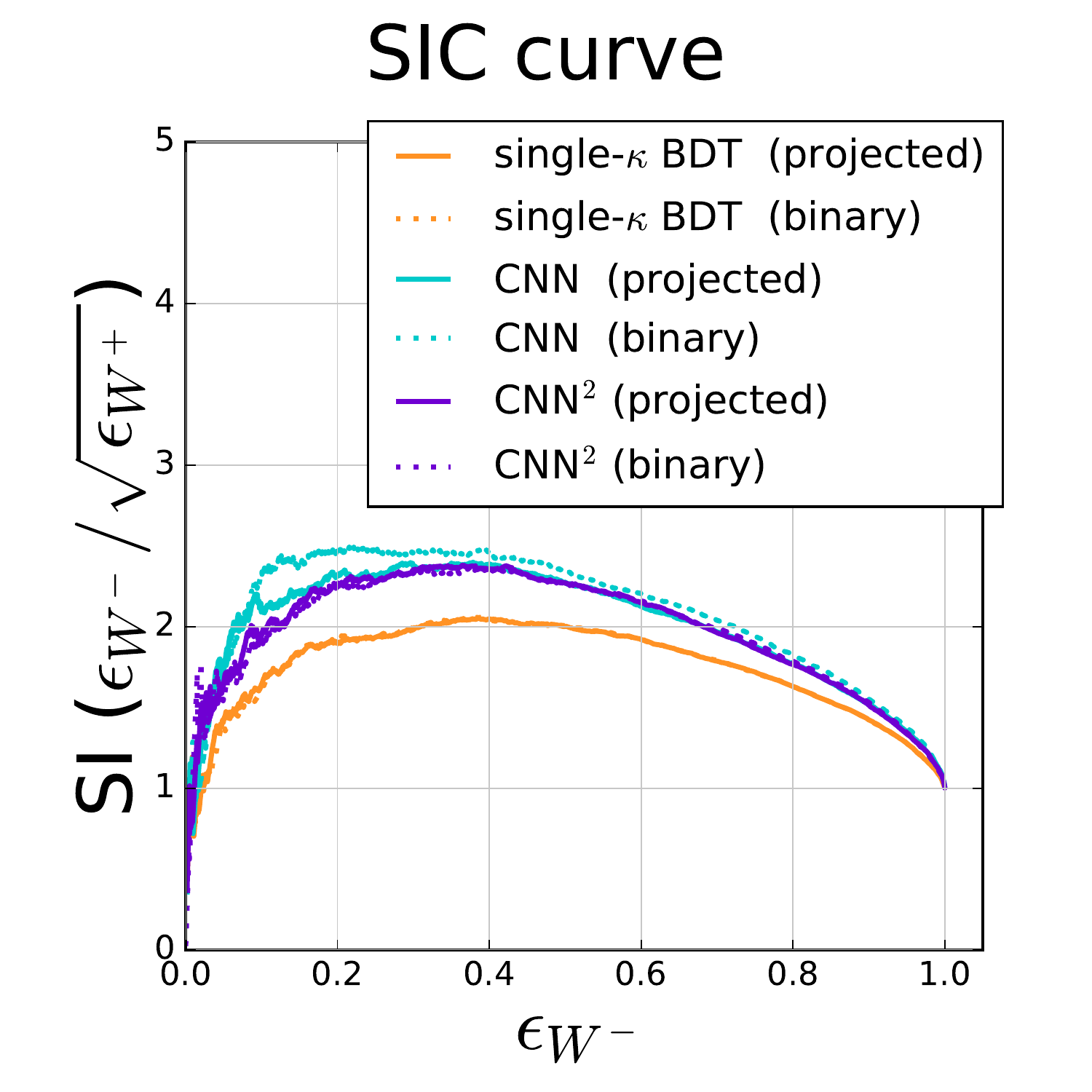}
\includegraphics[width=0.45\textwidth,angle=0]{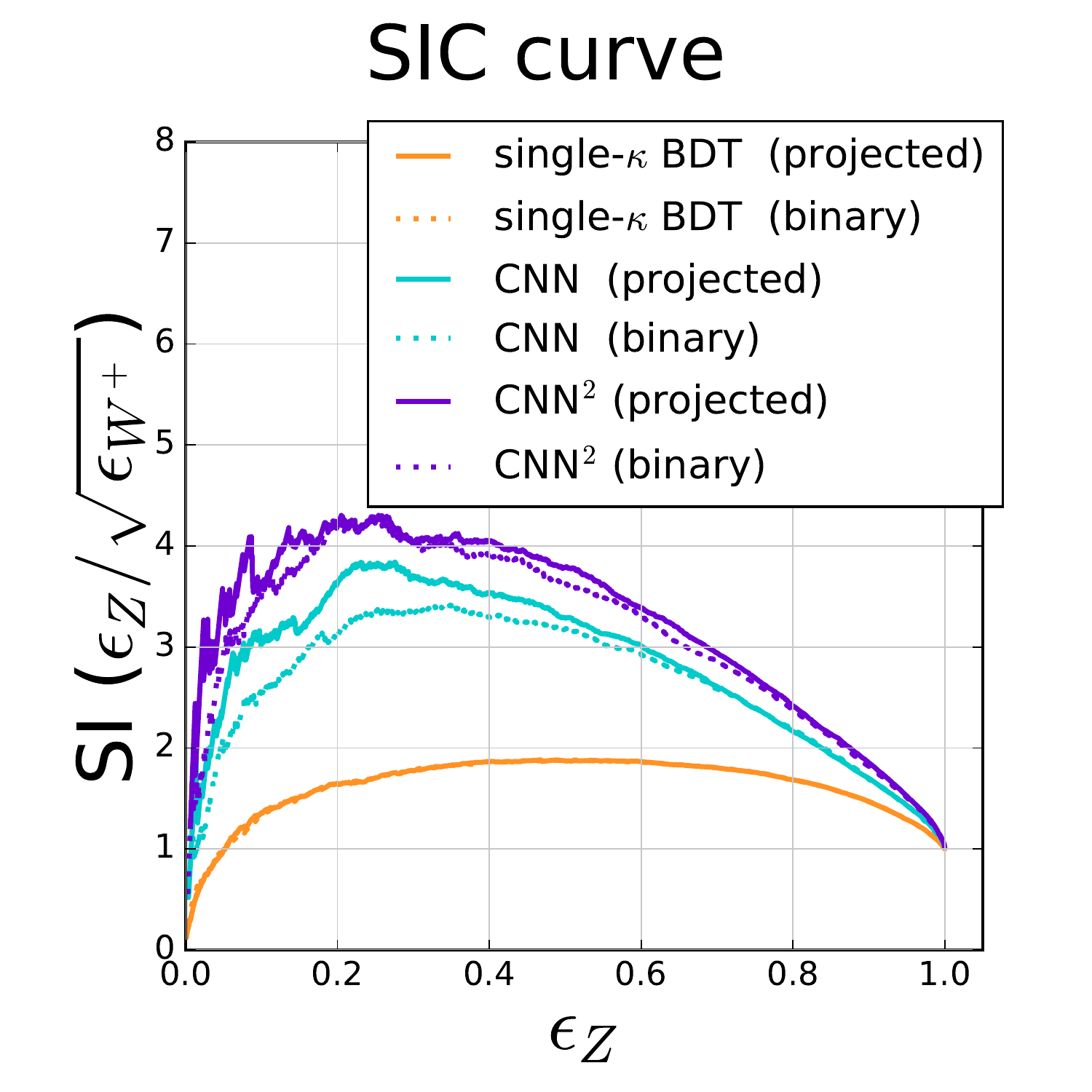}
\caption{SIC curves of the ternary classification for a $W^{-} / W^{+}$ discrimination (left) and for a $Z / W^{+}$ discrimination (right), when projected to binary according to eq.~\eqref{eq:probability}. The dashed curves are for binary classifications, and the solid curves for the projected ternary results.}
\label{fig:ternaryProjection}
\end{figure}

In figure~\ref{fig:ternaryProjection}, we plot the results of this projection in solid curves.  The dotted curves are the reproduced binary BDT and CNN results from figures~\ref{fig:800-SIC-Binary_wpwnFinal} and~\ref{fig:800-SIC-BinaryWZFinal} for comparison.  In the left plot, the solid and dotted curves are almost on top of each other. For the $Z / W^{+}$ projection, however, we observe that the ternary CNN outperforms the binary CNN after the projection in the low signal efficiency region. 

\section{What did the machine learn?} 
\label{sec:visualization}

In this section we will attempt to shed some light on how our CNN and CNN$^2$ taggers learn to classify $W^+/W^-/Z$ bosons, and the differences between them. Although a complete understanding is not possible -- they are still very much ``black boxes,'' we will find that with the help of some visualization techniques we can understand better what the machine has learned.

\subsection{Saliency maps}

Here we will use ``saliency maps" ~\cite{2013arXiv1312.6034S} to compare the CNN and CNN$^2$ networks in an attempt to understand why the latter outperforms the former. We will use the tool-kit of {\textsc{Keras-vis}}~\cite{raghakotkerasvis} to compute the saliency maps. Here the class saliency is extracted by computing the pixel-wise derivative of the class probability $P_{i}(x)$, as denoted in eq.~\eqref{eq:probability}:
\begin{equation}
w_i = \frac{\partial P_i(x)}{\partial x} 
~,
\label{eq:classmodelviz}
\end{equation}
where the gradient is obtained by back-propagation. By making a map of the gradients (\ref{eq:classmodelviz}) across an image, we can identify the regions of the image where the decision of the CNN (to be class $i$ or not) depends most sensitively.

\begin{figure}[H]
\centering
\includegraphics[width=0.85\textwidth,angle=0]{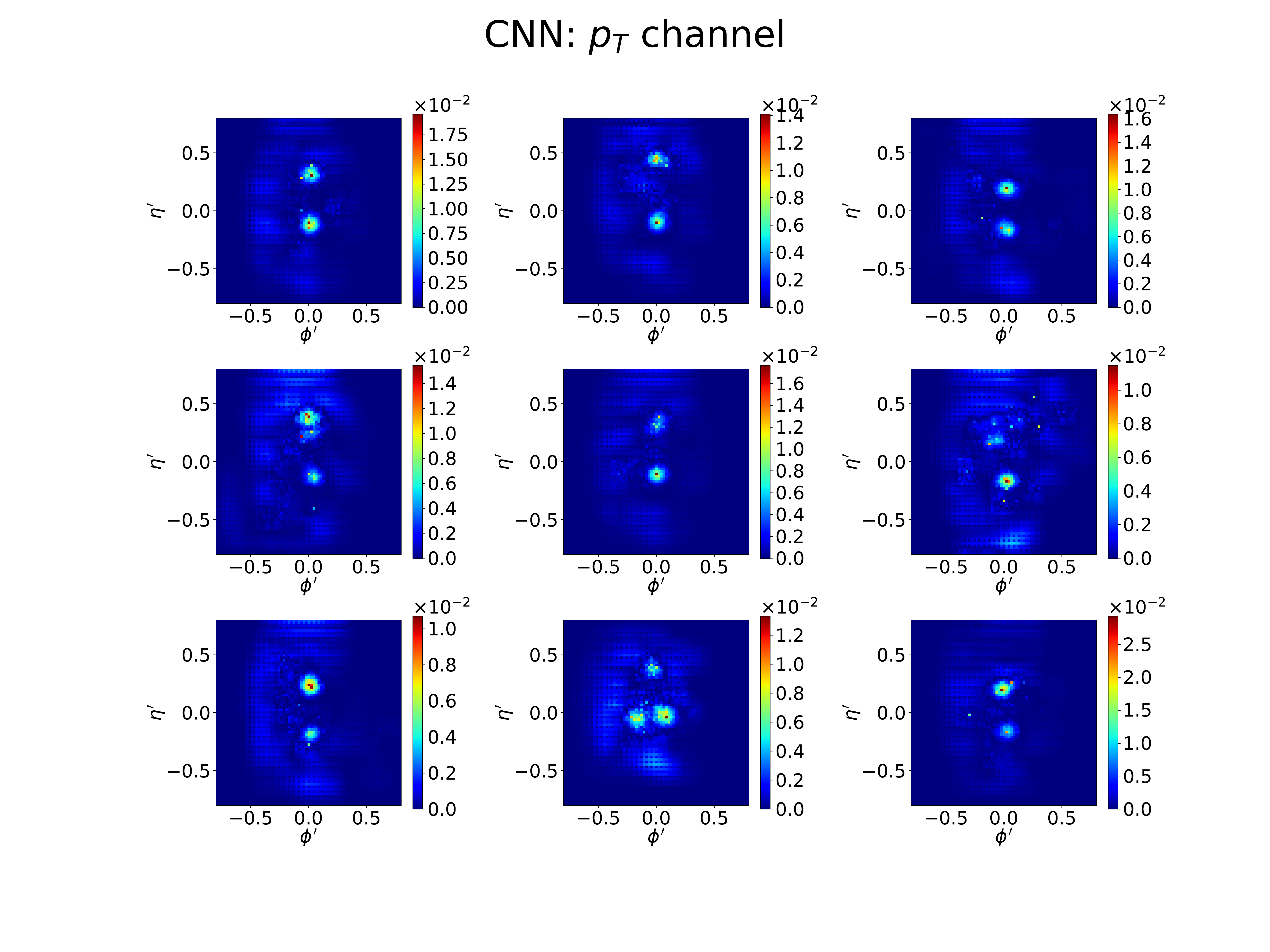}
\includegraphics[width=0.85\textwidth,angle=0]{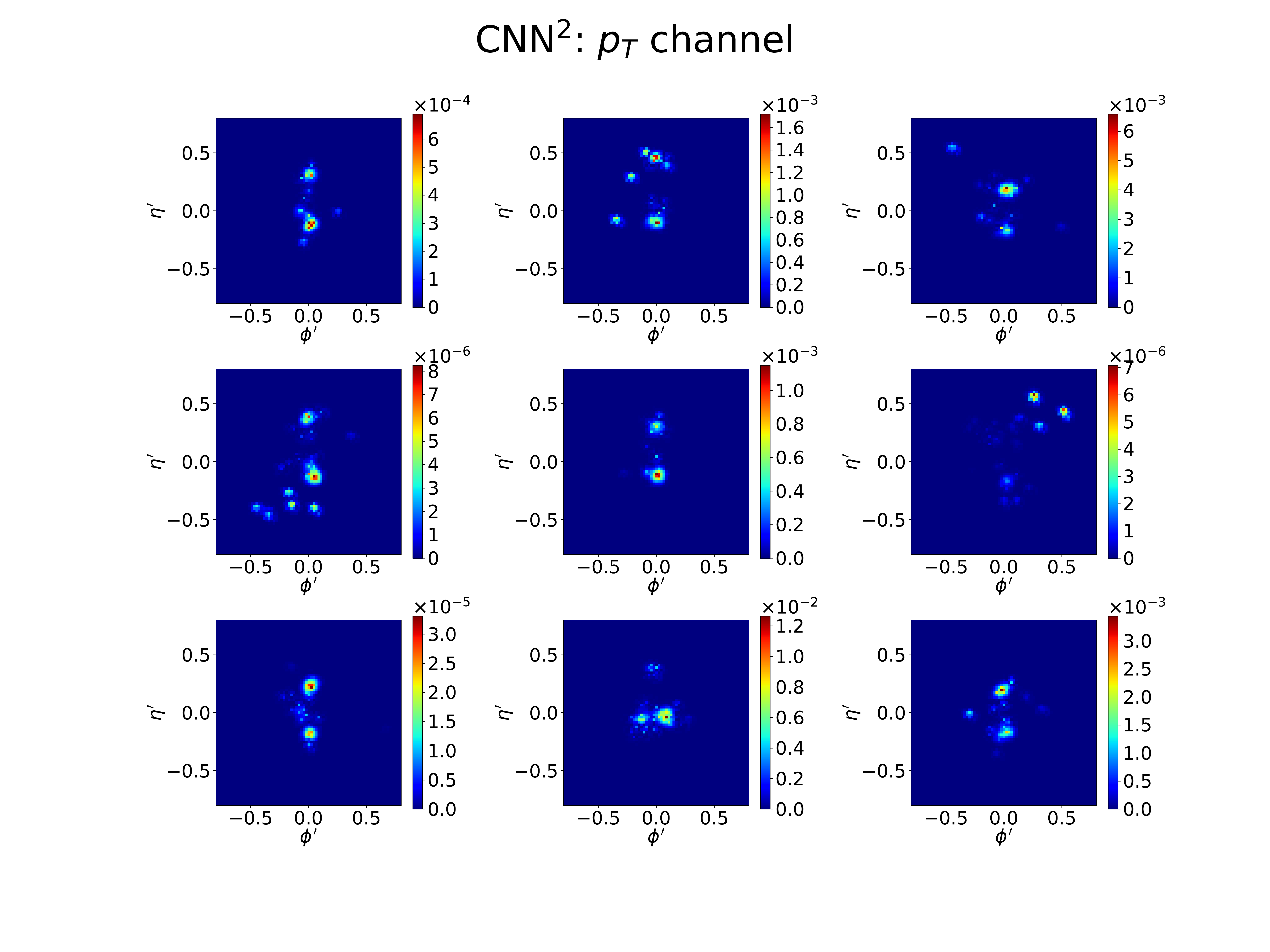}
\caption{Saliency maps for the $p_{T}$ channel of the CNN (upper plots) and CNN$^{2}$ (lower plots) networks on nine $W^-$ jet images from the test sample (for which both networks give correct output predictions). }
\label{fig:DeepShallowSaliencyPT}
\end{figure}

\begin{figure}[H]
\centering
\includegraphics[width=0.85\textwidth,angle=0]{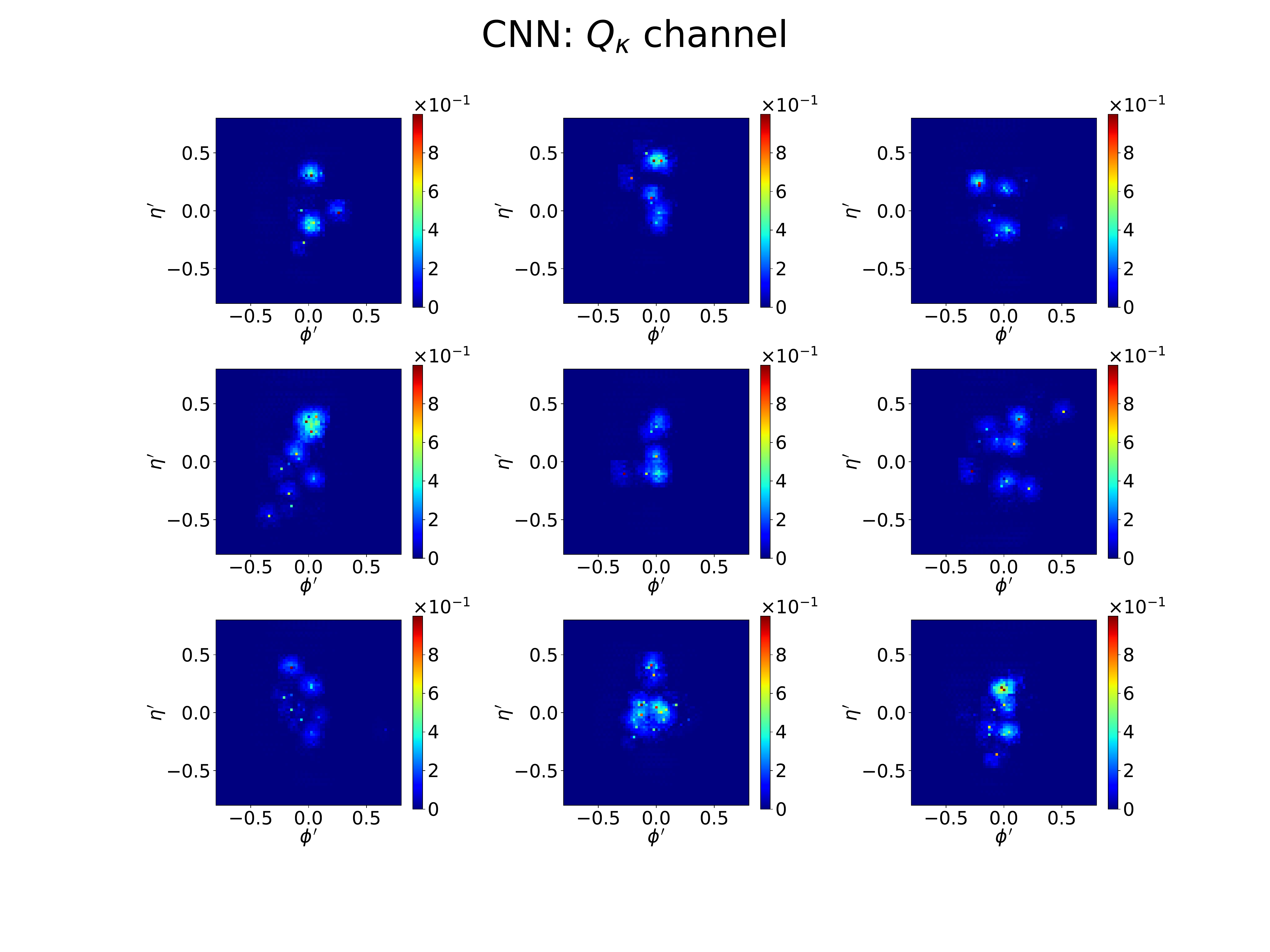}
\includegraphics[width=0.85\textwidth,angle=0]{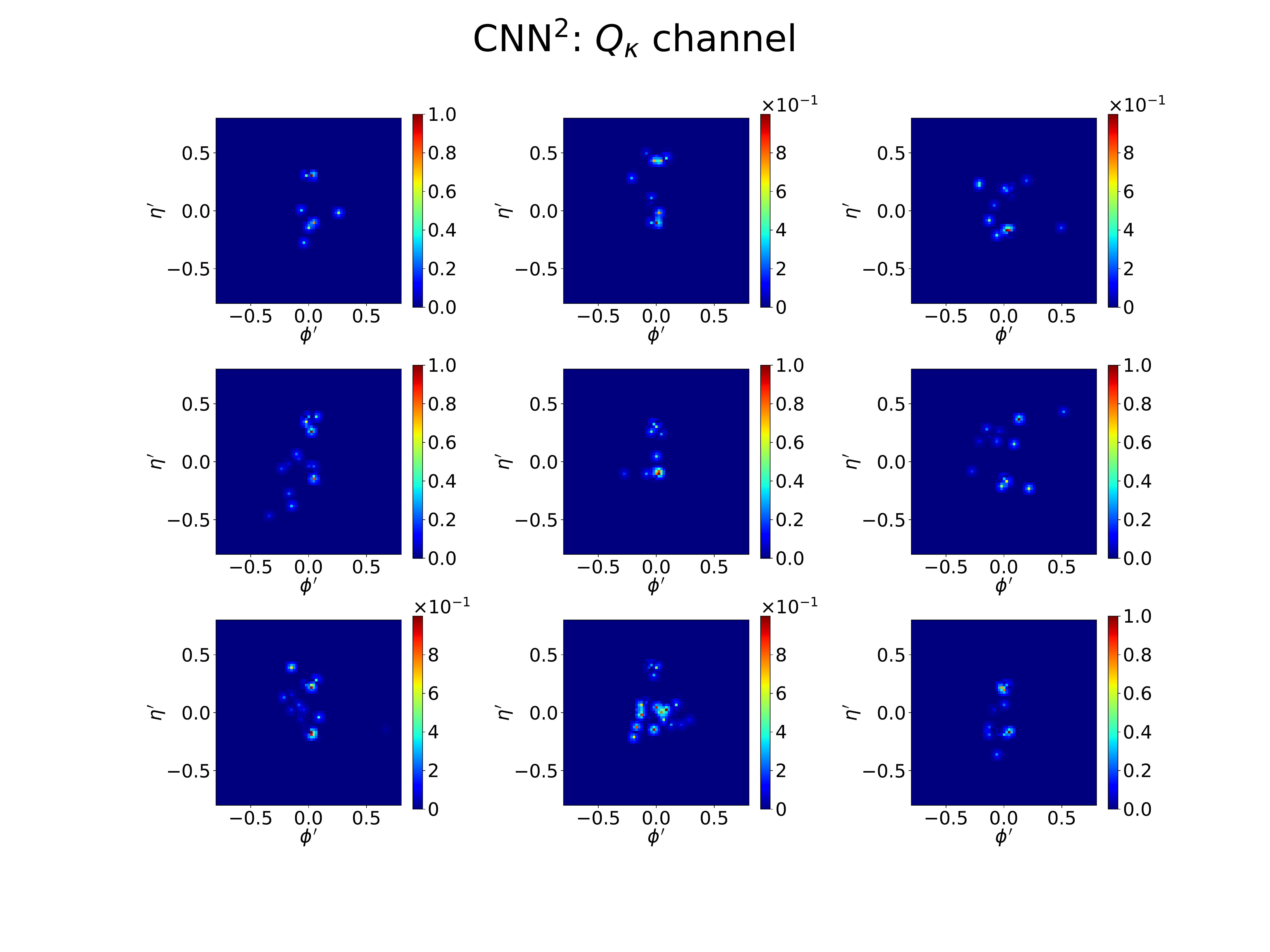}
\caption{Same as figure~\ref{fig:DeepShallowSaliencyPT} but for the $Q_\kappa$ channel. }
\label{fig:DeepShallowSaliencyQ}
\end{figure}

The saliency maps for nine $W^-$ jet images from the test sample are shown in figures~\ref{fig:DeepShallowSaliencyPT} (the $p_T$ channel) and \ref{fig:DeepShallowSaliencyQ} (the $Q_\kappa$ channel) for the CNN and the CNN$^2$ networks. The color in each pixel indicates the magnitude of the gradient value. 

The difference in the saliency maps between CNN and CNN$^2$ is very striking. We can see that the attention of the CNN$^{2}$ is generally concentrated on much smaller regions than the CNN. 
Evidently, the resolving power of the CNN$^2$ is much better than that of the CNN. Given that the 
CNN$^2$ network goes much deeper than the CNN network, this is perhaps expected: a deeper network structure with more convolutional layers is supposed to be more capable of capturing more subtle features in the training data. Altogether, this could explain why the CNN$^2$ mostly outperforms the CNN.

\pagebreak

\vfill

\vfill

\subsection{Phase transition in CNN's learning}

Another interesting difference between our networks is in their learning curves. We observe that the learning curve of the CNN always has a sudden jump in the performance. 
Such a {\it{phase transition}} in learning comes from the fact that the CNN tends to first learn characteristics of the $Z$ sample, and then those of the $W^+$ (or $W^-$) sample.

\begin{figure}[h]
\centering
\includegraphics[width=0.48\textwidth,angle=0]{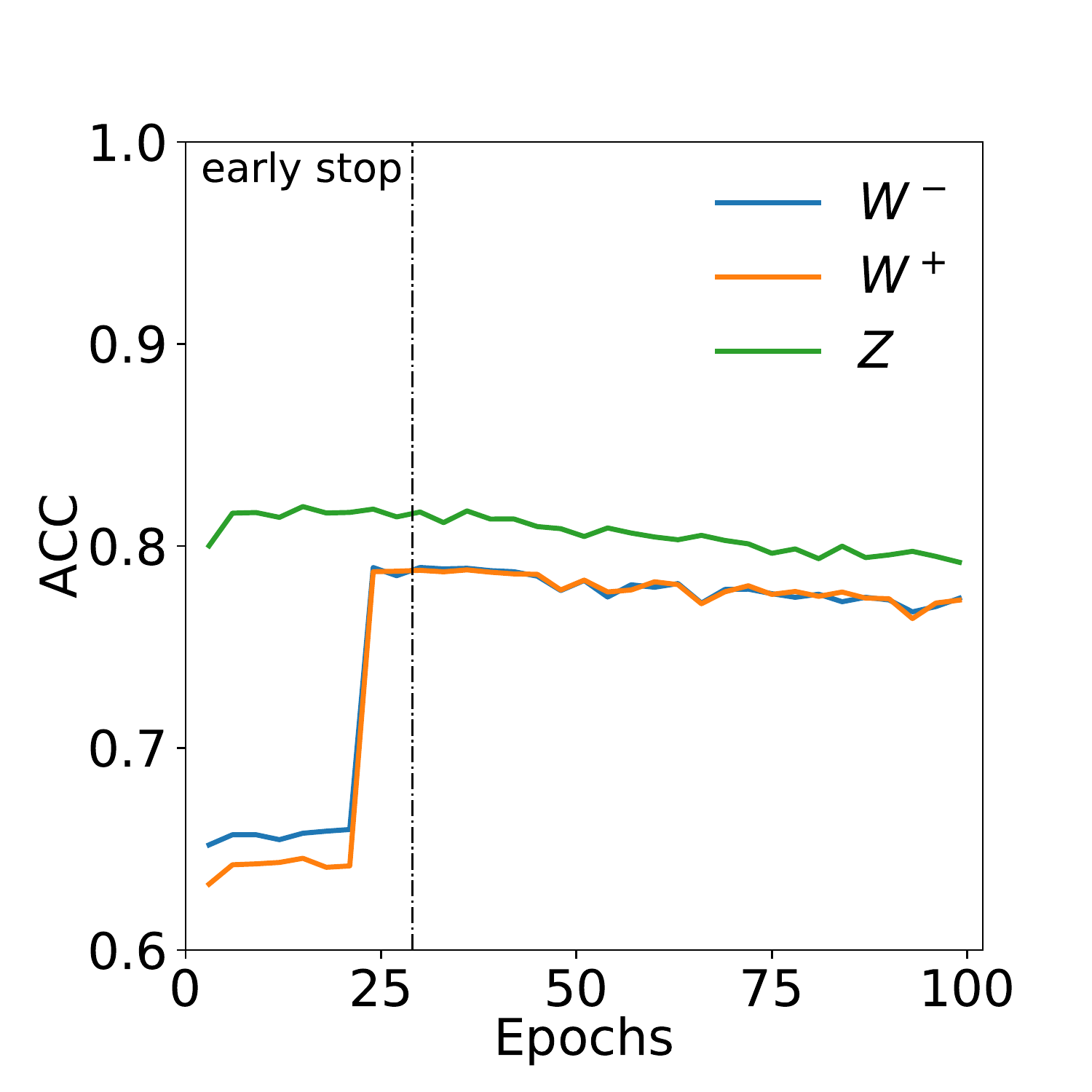}
\includegraphics[width=0.48\textwidth,angle=0]{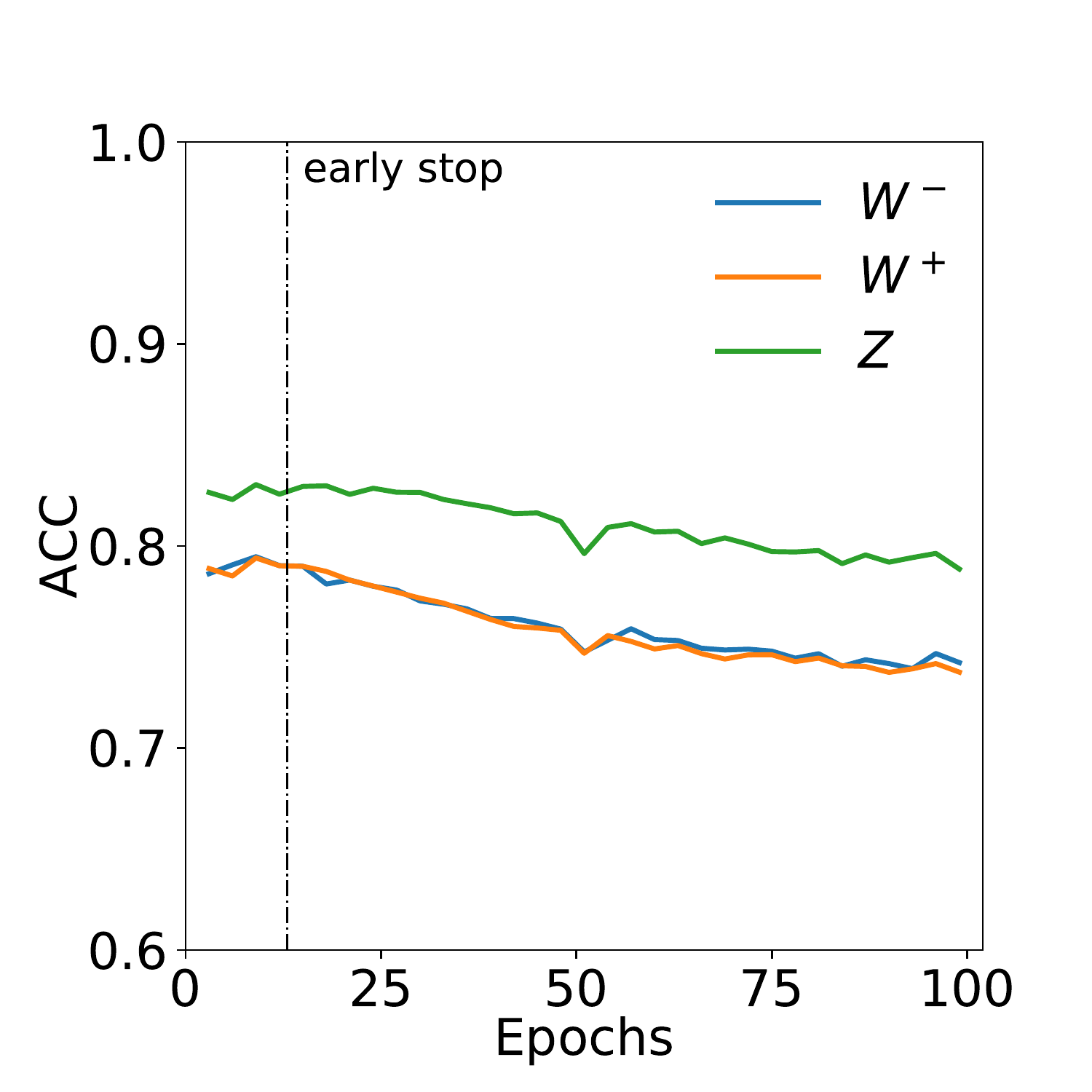}
\caption{Accuracy of a one-against-all metric, ACC, at different callback points for CNN (left) and CNN$^{2}$ (right). A phase transition in ACC during the CNN training stage occurs around the 25th epoch.}
\label{fig:PhaseTransition}
\end{figure}

In order to investigate this phase transition behavior, we monitor the network performance during its intermediate stages. During the training, we set a check point every three epochs and record the network's weights, then analyze to see how the discrimination ability evolves. At each check point, we evaluate the network performance using one-against-all metrics on the testing jet samples. The results of the ACC metric are shown in figure~\ref{fig:PhaseTransition}. We can see that the CNN develops the ability of $Z$-discrimination at an earlier stage. However, the network has not learned how to discriminate between $W^-$ and $W^+$ until after the 24th epoch, around which a phase transition is seen in their learning curves. 

Even though CNN$^{2}$ shows a more ``steady'' learning process, the possibility of phase transition phenomenon cannot be ruled out. It is possible that the CNN$^{2}$ learns so fast that the performance in all classes saturates within one epoch and, therefore, the phase transition is not manifest.

\section{Conclusions}
\label{sec:summary}

In this work, we apply modern deep learning techniques to build better taggers of boosted, hadronically-decaying weak gauge bosons. We demonstrate and provide the results for the boosted weak bosons with $p_T \sim 400$~GeV throughout this paper. (We have also studied the scenario with even more boosted $W$ and $Z$ bosons, with $p_T \sim 1$~TeV, and found results similar to what are presented here.) Going beyond previous works, we incorporate jet charge information
in order to discriminate between positively and negatively charged $W$ bosons, and between $W$ and $Z$ bosons. We study all possible binary classification tasks as well as the full ternary classification problem. Taking BDT and cut-based taggers as our baselines for comparison, we construct a simple CNN tagger that takes jet images as the input, and show that it leads to significant gains in classification accuracy and background rejection.  

In addition to the simple CNN tagger, we also construct a novel CNN structure consisting of two parallel CNNs, which we call CNN$^2$. The key feature of this structure is to assign different network depths to each of the $p_{T}$ and $\mathcal{Q}_{\kappa}$ channels. This further improves the performance of nearly all the classification tasks.

We see various ways in which our work could be extended and improved.  First,
traditional CNNs may have some drawbacks due to the fact that the receptive field of every neuron, which is the field of view that one unit can perceive, is fixed by the assigned kernel sizes and depth of the network. But the complexity in detecting the patterns or features in realistic problems generally differ. As the $W^{-}/W^{+}/Z$ discrimination problem in our analysis shows, the complexity/depth for dealing with $W^{-}/W^{+}$ and $W/Z$ is different. To optimize the performance as well as the computational costs, a ``ResNet'' network architecture with ``skip connections"~\cite{Kasieczka:2019dbj,DBLP:journals/corr/HeZR016,DBLP:journals/corr/LuoLUZ17} may be a desirable solution. It would be interesting to study this further, along with other architectures and jet representations such as point clouds and sequences. 

Another direction where it may be possible to improve on this work is to come up with an architecture that takes $p_T$ and charge information as totally separate channels, such that the network could learn the ideal combination of them for measuring the charge of the boosted heavy resonance. The fact that our tagger performance depends on the value of $\kappa$ is a symptom that it is not truly learning the optimal combination of $p_T$ and charge information.

Comparing the performance gains due to deep learning seen in this work with recent related works in the literature~\cite{Aad:2015eax,Fraser:2018ieu}, we have seen similar improvements over more conventional methods. We caution that these are merely rough comparisons, as the nature and details of the classification problems are different from ours. Nevertheless, this gives further evidence for the enormous potential of deep learning for the study of jet substructure and boosted resonance tagging.

\section*{Acknowledgments}

We gratefully acknowledge the support of NVIDIA Corporation with the donation of the Titan Xp GPU used for this research. We thank Gregor Kasieczka and Ben Nachman for helpful discussions. Y.-C. C. thanks Yi-Ting Lee for the advice on the technical details in deep learning. C.-W. C. was supported in part by the Ministry of Science and Technology (MOST) of Taiwan under Grant No. MOST-104-2628-M-002-014-MY4. G.C. acknowledges support by Grant No. MOST-107-2811-M-002-3120 and CONICYT-Chile FONDECYT Grant No. 3190051. The work of D.S. was supported by the US Department of Energy 
under grant DE-SC0010008.

\appendix
\section{Check of soft sensitivity using power showering from HERWIG}
\label{sec:herwig}

In this appendix, we show the performance of our taggers on a separate set of jet samples using a different parton shower model.  This is to study whether our deep learning jet-tagging method is vulnerable to the modeling of parton showing and hadronization. For this dataset, the parton-level hard process events generated by \textsc{MadGraph5}, the detector simulation carried out by \textsc{DELPHES} and the clustering procedure by \textsc{FastJet} are controlled to be identical to those introduced in Sec.~\ref{sec:observables}.  The only difference is that here we use \textsc{HERWIG 7.1.5}~\cite{Bahr:2008pv, Bellm:2015jjp} for showering and hadronization. Detailed tuning is referred to the CMS software collection (CMSSW).

For the quality cuts, the same selection criteria (table~\ref{tab:selections}) are performed on the raw jet data. The number of jets for later training and testing purpose are listed in table~\ref{tab:JetSamplesHerwig}. 
And overall, the jet variables and jet images of this Herwig-jet sample have been checked to be consistent with those obtained from the Pythia-jet sample.

\begin{table}[h]
\begin{center}
\begin{tabular*}{0.75\textwidth}{c|c|c}
\bottomrule
 & \hspace{0.2cm} Training set\hspace{0.2cm}  & \hspace{0.2cm} Testing set \hspace{0.2cm} \\
\hline 
\hspace{0.2cm} Jet sample size    & \hspace{0.2cm} $191k+201k+176k$ \hspace{0.2cm} & $38k+40k+35k$ \hspace{0.2cm}\\
\toprule
\end{tabular*} 
\caption{Summary of the Herwig-jet sample sizes used for training and testing, after the selections in table~\ref{tab:selections}. The entries in the sums correspond to the ($W^+, W^-, Z$) samples, respectively. In the training stage, the training set is also divided into two subsets as described in table~\ref{tab:JetSamples}.}
\label{tab:JetSamplesHerwig}
\end{center}
\end{table}

\begin{figure}[h]
\centering
\includegraphics[width=0.45\textwidth,angle=0]{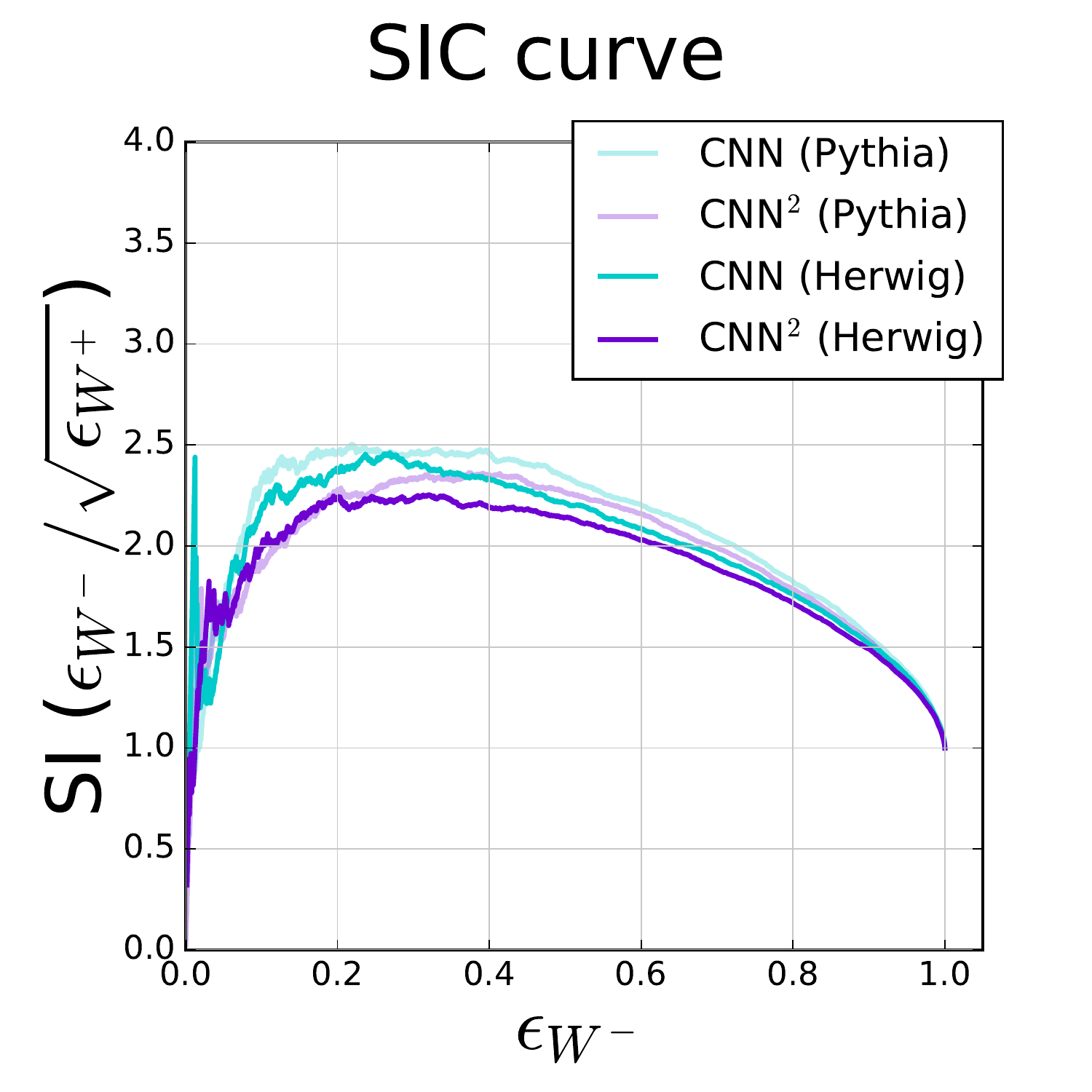}
\includegraphics[width=0.45\textwidth,angle=0]{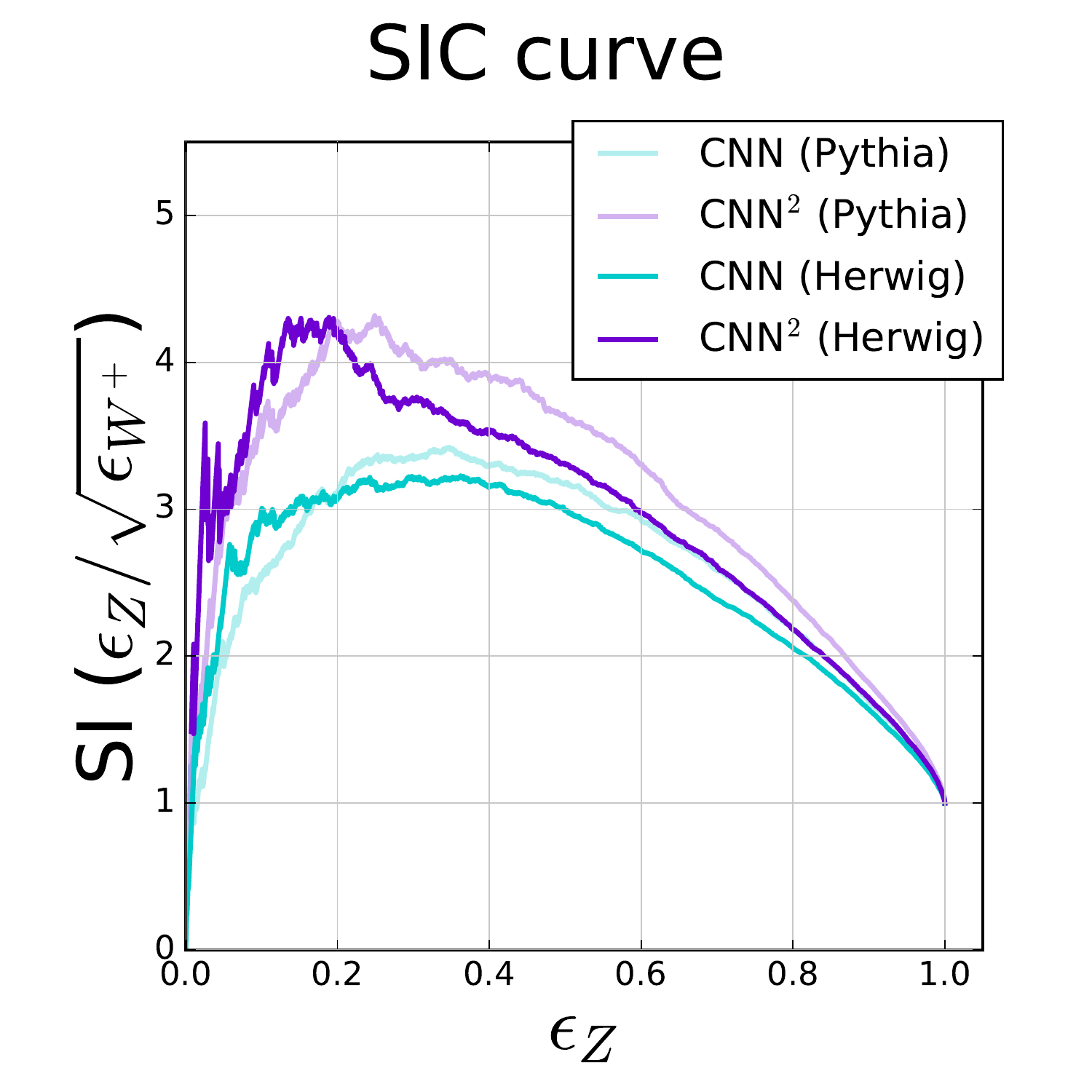}
\caption{SIC curves for the binary classifications to discriminate $W^{-}$ from $W^{+}$ (left) and $Z$ from $W^{+}$ (right) using the CNN and CNN$ ^2$ taggers.}
\label{fig:800-SIC-Binary_wpwnFinalHerwig}
\end{figure}

\begin{figure}[h]
\centering
\includegraphics[width=0.45\textwidth,angle=0]{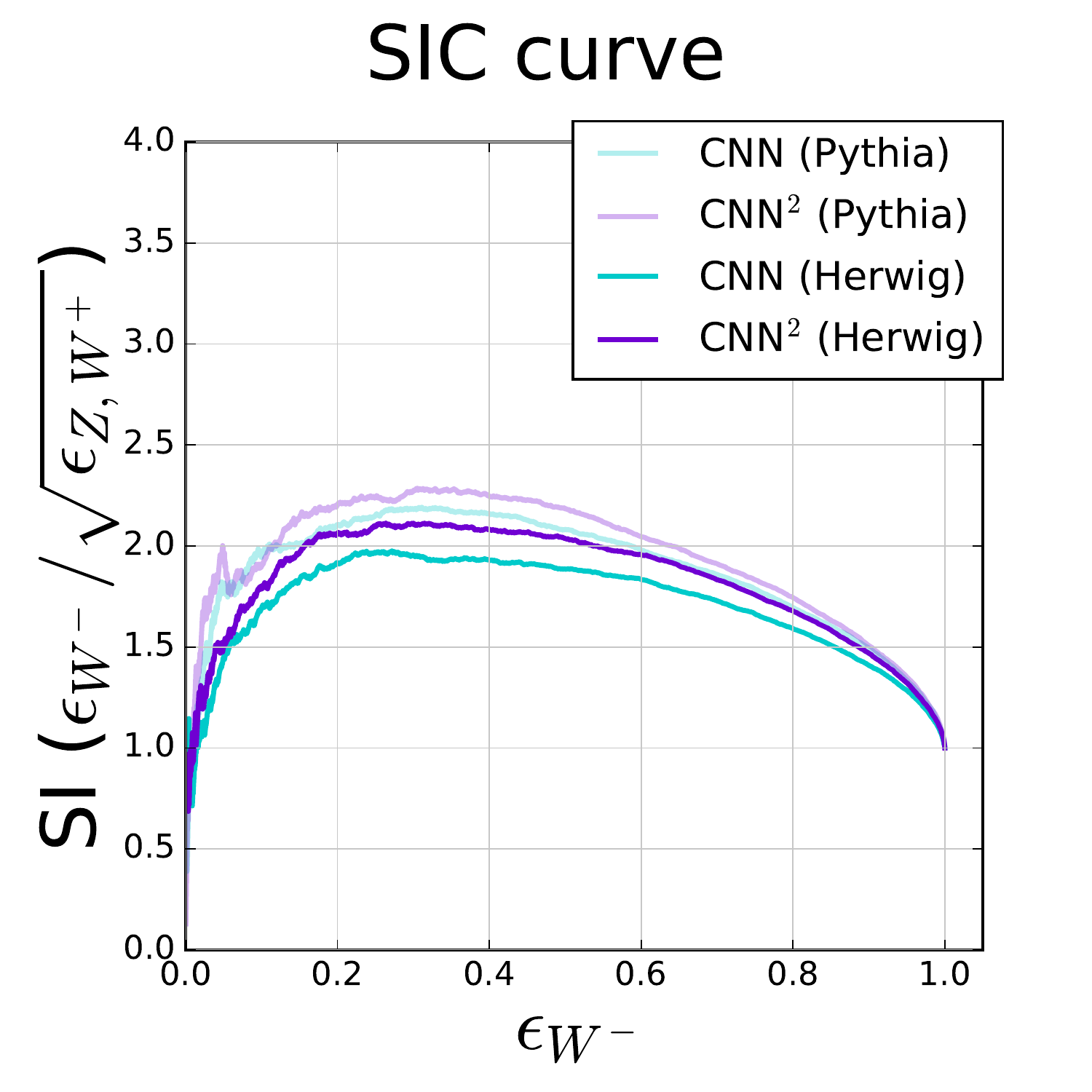}
\includegraphics[width=0.45\textwidth,angle=0]{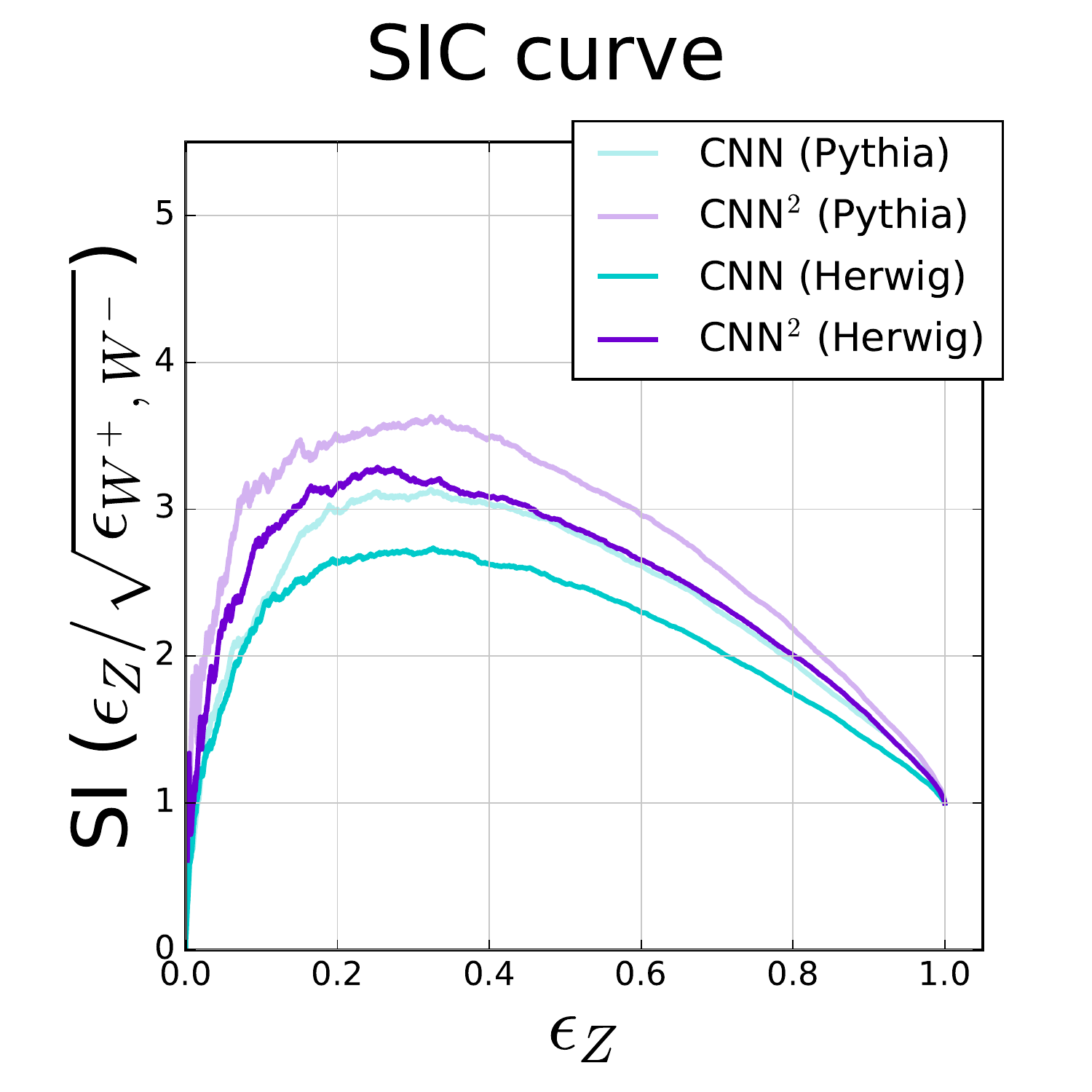}
\caption{SIC curves for the ternary classifications to discriminate $W^{-}$ from the rest (left) and $Z$ from the rest (right) using the CNN and CNN$ ^2$ taggers. }
\label{fig:800-SIC-TernaryFinalHerwig}
\end{figure}

In the following, the results of CNN and CNN$^{2}$, whose structures are described in table~\ref{tab:Networks}, working on the Herwig-jet sample for binary $W^{-} / W^{+}$, $Z / W^{+}$ and ternary classifications are demonstrated. For figures~\ref{fig:800-SIC-Binary_wpwnFinalHerwig} and ~\ref{fig:800-SIC-TernaryFinalHerwig}, the SIC curves for the Herwig dataset (darker colors) are overlapped on top of the Pythia results (lighter colors), which are already shown in figures~\ref{fig:800-SIC-Binary_wpwnFinal}, ~\ref{fig:800-SIC-BinaryWZFinal} and ~\ref{fig:800-wp-SIC-TernaryFinal}, ~\ref{fig:800-z-SIC-TernaryFinal}. And the corresponding performance metrics are provided in tables~\ref{tab:CNNw_binaryHerwig} and ~\ref{tab:800-SIC-TernaryHerwig}.
The results are slightly worse than those from the Pythia-jet dataset, based upon which all the taggers in this work are optimized.  Nevertheless, the tagger performance indicated by the SIC curves roughly agrees with what is observed in Pythia-jet dataset. This shows that the tagging abilities of our CNN and CNN$^{2}$ taggers are independent of showering and hadronization models employed in the analysis.

\begin{table}[h]
\begin{center}
\begin{tabular*}{0.85\textwidth}{c|ccc|ccc}
\bottomrule
& \multicolumn{3}{c|}{$W^{-} / W^{+}$} & \multicolumn{3}{c}{$Z / W^{+}$} \\
& \hspace{0.2cm}   R50 &\hspace{0.25cm}   AUC\hspace{0.25cm}  &\hspace{0.25cm}   ACC \hspace{0.25cm} 
& \hspace{0.2cm}   R50 &\hspace{0.25cm}   AUC\hspace{0.25cm}  &\hspace{0.25cm}   ACC \hspace{0.25cm} 
\\ \hline
CNN      	   &  19.5819  &  0.8788  &  0.7970  &  35.7788  &  0.9013  &  0.8253  \\
CNN$ ^2$ 	   &  18.3411  &  0.8728  &  0.7920  &  43.6485  &  0.9109  &  0.8336  \\
\toprule 
\end{tabular*} 
\caption{Performance metrics for CNN and CNN$ ^2$ taggers on the Herwig-jet sample in the $W^-/W^+$ and $Z / W^{+}$ binary classification tasks.}
\label{tab:CNNw_binaryHerwig}
\end{center}
\end{table}

\begin{table}[h]
\begin{center}
\begin{tabular*}{0.83\textwidth}{c|c|ccc|ccc}
\bottomrule
  &  overall   &   \multicolumn{3}{c|}{signal: $W^{-}$} 
  &  \multicolumn{3}{c}{signal: $Z$} 
\\
           &  \hspace{0.1cm} ACC \hspace{0.1cm}
           &  \hspace{0.02cm}  R50 &\hspace{0.02cm}  AUC\hspace{0.02cm}  &\hspace{0.02cm}  ACC \hspace{0.02cm} 
           &  \hspace{0.02cm}  R50 &\hspace{0.02cm}  AUC\hspace{0.02cm}  &\hspace{0.02cm}  ACC \hspace{0.02cm} 	\\
           \hline
CNN	      & 0.6943 & 14.2221 &  0.8556  &  0.7742 & 24.8287 &  0.8734  &  0.7980\\
CNN$^2$      & 0.7197 & 16.6117 &  0.8677  &  0.7865 & 33.6355 &  0.8961  &  0.8212\\
\toprule 
\end{tabular*} 
\caption{Performance metrics for CNN and CNN$ ^2$ taggers on the Herwig-jet sample in the ternary classification tasks.}
\label{tab:800-SIC-TernaryHerwig}
\end{center}
\end{table}

\bibliographystyle{JHEP}
\bibliography{jetWcharge}

\end{document}